\documentclass[sigconf, nonacm]{acmart}

\pdfoutput=1

\usepackage{subcaption}
\usepackage{hyphenat} % use \hyp{}
\usepackage[normalem]{ulem} % to use \sout for strikethrough
\usepackage{color}
\usepackage{xcolor}
\usepackage{graphicx}
\usepackage{mathtools}
\usepackage{soul}
\usepackage[linesnumbered,ruled,vlined]{algorithm2e}
\usepackage{bm}
\usepackage[LGR,T1]{fontenc} % notice LGRx instead of LGR
\usepackage[utf8]{inputenc}   % utf8 is required
\usepackage{multirow}

\newtheorem{definition}{Definition}
\newtheorem{problem}{Problem}
\newcommand{\thiswork}{A\textsc{trapos}\xspace} % name of this work
\newcommand{\dblp}{{Scholarly HIN}\xspace} 
\newcommand{\gdelt}{{News articles HIN}\xspace}
\newcommand{\textgreek}[1]{\begingroup\fontencoding{LGR}\selectfont#1\endgroup}

\begin{document}

\title{\thiswork: Evaluating Metapath Query Workloads in Real Time}

\author{Serafeim Chatzopoulos}
\orcid{0000-0003-1714-5225}
\affiliation{%
  \institution{Univ. of the Peloponnese \& \\IMSI, ``Athena'' RC}
  \country{Greece}
}
\email{schatzop@uop.gr}

\author{Thanasis Vergoulis}
\affiliation{%
  \institution{IMSI, ``Athena'' RC}
    \country{Greece}
}
\email{vergoulis@athenarc.gr}

\author{Dimitrios Skoutas}
\affiliation{%
  \institution{IMSI, ``Athena'' RC}
  \country{Greece}
}
\email{dskoutas@athenarc.gr}

\author{Theodore Dalamagas}
\affiliation{%
  \institution{IMSI, ``Athena'' RC}
  \country{Greece}
}
\email{dalamag@athenarc.gr}

\author{Christos Tryfonopoulos}
\affiliation{%
  \institution{Univ. of the Peloponnese}  
  \country{Greece}
}
\email{trifon@uop.gr}

\author{Panagiotis Karras}
\affiliation{%
  \institution{Aarhus University}
  \country{Denmark}
}
\email{piekarras@gmail.com}

\renewcommand{\shortauthors}{Chatzopoulos and Vergoulis, et al.}

\begin{abstract}
\emph{Heterogeneous information networks} (HINs)
represent different types of entities and relationships between them. Exploring, analy\-sing, and extracting knowledge from such networks relies on \emph{metapath queries} that identify pairs of entities connected by relationships of diverse semantics. While the real-time evaluation of metapath query workloads on large, web-scale HINs is highly demanding in computational cost, current approaches do not exploit interrelationships among the queries. In this paper, we present \thiswork, a new approach for the real-time evaluation of metapath query workloads that leverages a combination of efficient sparse matrix multiplication and intermediate result caching. \thiswork selects intermediate results to cache and reuse by detecting frequent sub-metapaths among workload queries in real time, using a tailor-made data structure, the \emph{Overlap Tree}, and an associated caching policy. Our experimental study on real data shows that \thiswork~accelerates exploratory data analysis and mining on HINs, outperforming off-the-shelf caching approaches and state-of-the-art research prototypes in all examined scenarios.
\end{abstract}

\maketitle

\section{Introduction}\label{sec:intro}

\emph{Heterogeneous information networks (HINs)}, also known as \emph{labeled property graphs}, offer an intuitive and generic model to encapsulate complex semantic information via different types of nodes and edges~\cite{HINsurvey}. By this model, each node (or edge) has internal structure, including a set of \emph{properties}. By virture of the compactness of this model compared to those of other types of \emph{Knowledge Graphs} like \emph{RDF triple stores}~\cite{rdf3x,hexastore}, HINs are increasingly supported by data systems, e.g., Neo4j\footnote{\url{https://neo4j.com/}}~\cite{neo4jfund}, and investigated in research studies~\cite{fang2021cohesive, shen2014probabilistic, fang2020effective, jian2020effective, sun2012relation, sun2012mining, sun11, chen2021structure, li2021leveraging, xie2021sequential, wang2020howsim, yang2020effective}.

\begin{figure}[!t]
\centering
\includegraphics[width=0.98\linewidth]{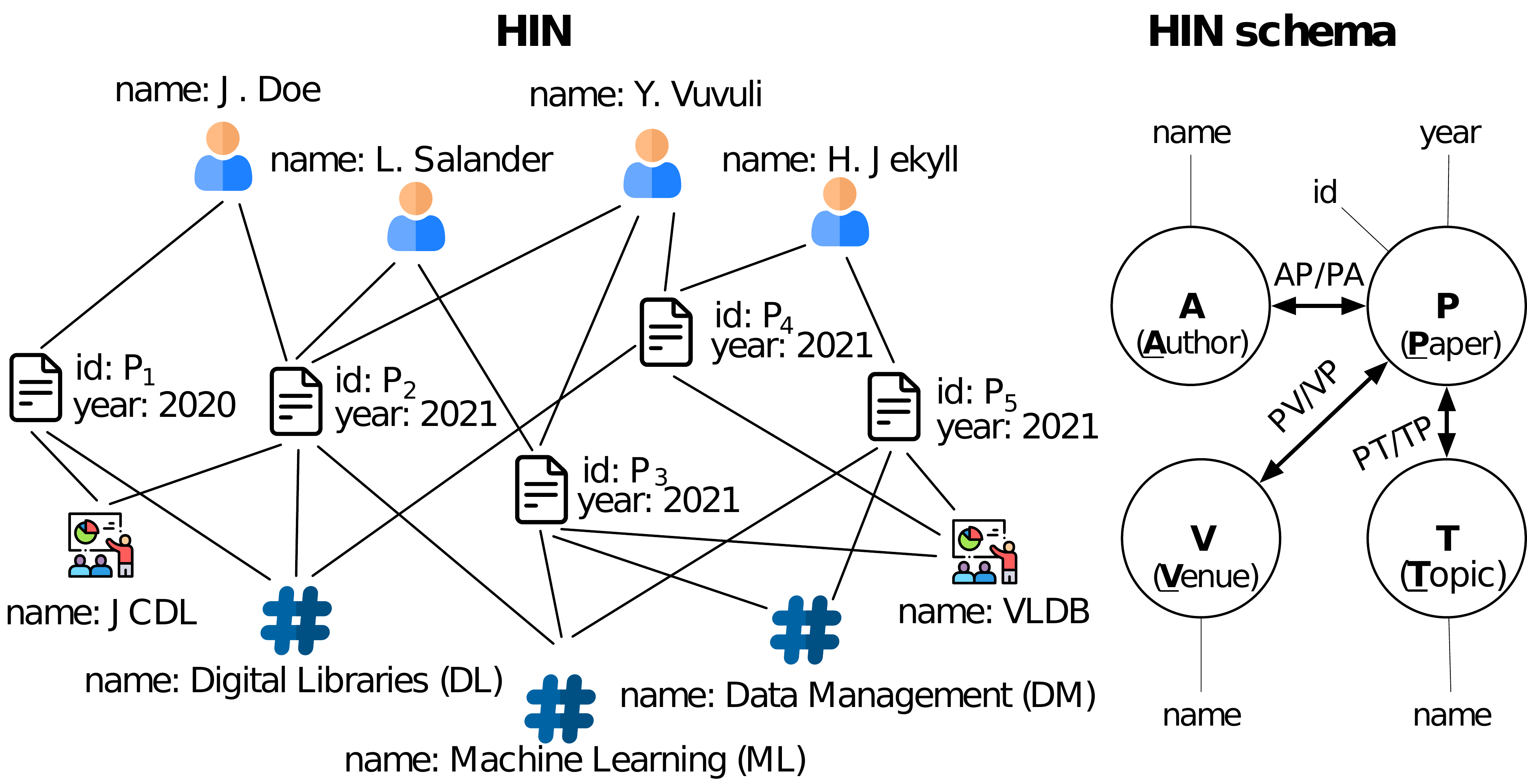}
\vspace{-3mm}
\caption{An example HIN and its implied schema.}\label{fig:hin-example}
\end{figure}

An example HIN capturing scholarly data, along with its implied \emph{schema}\footnote{It is worth mentioning that HINs are naturally schemaless, i.e., there is not any technical requirement for a schema to be predetermined. However, the current `schema' can be implied from the data and it can be conceptually useful in some cases (e.g., this is why Neo4j gives this option with the \texttt{call db.schema()} command.} that represents the involved entity types, relationship types, and properties, is illustrated in Figure~\ref{fig:hin-example}.
It consists of nodes representing papers (\texttt{P}), authors (\texttt{A}), venues~(\texttt{V}), and topics (\texttt{T}) and (bidirectional) edges of three types: authors -- papers (\texttt{AP} / \texttt{PA}), papers -- topics (\texttt{PT} / \texttt{TP}), and papers -- venues (\texttt{PV} / \texttt{VP}).

Indirect relationships between entities are implicitly encoded by paths in the HIN. In particular, all paths that correspond to the same sequence of node and edge types (i.e., the same \emph{`metapath'}~\cite{sun11}) encode latent relationships of the same interpretation between the starting and ending nodes. For example, in the HIN of Figure~\ref{fig:hin-example}, the metapath $\langle A P T P A \rangle$  relates authors that have published papers on the same topic (e.g., both \texttt{`J. Doe'} and \texttt{`H. Jekyll'} have papers about \texttt{`DL'}). It is also worth mentioning that each metapath corresponds to a path on the schema of the HIN.

Metapaths are instrumental for HIN exploration and analysis from multiple perspectives, as a means to derive different views of such a network. For example, the metapath-based connectivity can be used to define node similarity measures~\cite{sun11, Hetesim, shi2012, joinsim} or to rank nodes based on their centrality in a metapath-defined network~\cite{hrank, multirank, li2014, pathrank}. In the previous example, using the \texttt{TPT} metapath to perform a similarity join could reveal that topics \texttt{`ML'} and \texttt{`DL'} are very related since they are involved in two common papers. Moreover, to further elaborate the analysis, it is often useful to apply property-based \emph{constraints} to a given metapath (e.g., in the previous example, to consider only metapath instances involving papers published later than \texttt{`2020'}). Motivated by the above, we have recently developed prototypes for supporting metapath-based exploration and mining of HINs~\cite{chatzopoulos2020sphinx, chatzopoulos2021scinem}.

The first step in any metapath-based analysis is to generate a transformed network that contains edges between all pairs of nodes connected with one or more paths of the specified type. This preprocessing step is computationally expensive. While various methods aim to speed up the computation of a \emph{single} metapath query, which typically involves a series of matrix multiplications~\cite{hrank}, the problem becomes even more severe in scenarios calling for the real-time evaluation of multiple metapath queries. In metapath-based feature selection~\cite{DBLP:conf/icdm/ShiW14, DBLP:conf/www/MengCMSZ15}, different metapaths reveal different associations among entities, informing tasks like recommendation and link prediction~\cite{DBLP:conf/icdm/CaoKY14, DBLP:conf/kdd/ChenYWWNL18, DBLP:journals/tkde/ShiHZY19}. Such a task requires a large number of metapath queries (i.e., a \emph{query workload}), leading to a significant bottleneck in the data analysis process. Yet, in such a multi-query evaluation scenario, significant overlaps typically occur among metapath queries~\cite{finkelstein1982common, sellis1988multiple, sellis1990multiple, roy2000efficient, kathuria2017efficient}; such overlaps are translated into repetitive matrix multiplications. We may thus avoid a large portion of the heavy computations involved using query materialization; yet, to the best of our knowledge, no existing approach does so.

In this paper, we introduce \thiswork,\footnote{From Greek \textgreek{>atrap'os}, `path', what is frequently trodden.} a metapath query evaluation approach that detects, in real time, frequent metapath overlaps within a sequence of queries. To do so, it uses: (i) sparse matrix representations, which leverage the fact that the matrices involved in metapath computations are largely sparse, especially for constrained metapaths; and (ii) a tailored data structure with a customised caching policy to materialise reoccurring intermediate results. Our experimental evaluation demonstrates that \thiswork~ outperforms traditional, single-query approaches in the evaluation of metapath query workloads, while it is considerably faster than baseline approaches that we implemented. To the best of our knowledge, this work is the first to study the problem of optimizing the real-time evaluation of metapath query workloads. Our contributions can be summarised as follows:

\begin{itemize}
    \item We show how metapath computations can be accelerated using matrix multiplication algorithms tailored for sparse matrices along with an appropriate cost model, especially when dealing with constrained metapaths.
    \item We introduce a new data structure, the \emph{Overlap Tree}, that reveals overlaps among metapaths and a cache insertion policy using this structure to cache and reuse intermediate results in real-time while running a metapath query workload.
    \item We propose a tailored, effective cache replacement policy that exploits cache item interdependence.
    \item We conduct a thorough evaluation on real data showcasing the efficiency of our approach.
\end{itemize}

The rest of the paper is structured as follows. In Section~\ref{sec:bkgr}, we formally define the problem. Section~\ref{sec:approach} introduces our approach, \thiswork. In Section~\ref{sec:evaluation}, we discuss our experimental evaluation. Section~\ref{sec:related} reviews related work and Section~\ref{sec:conclusions} concludes the paper.

\section{Problem Definition}\label{sec:bkgr}

Here, we introduce the basic concepts and notations we use, and define the problem of \emph{metapath query workload evaluation} (MQWE).

A \emph{heterogeneous information network (HIN)}~\cite{HINsurvey,sun11}, also known as knowledge graph, is a directed multigraph that contains multiple types (or \emph{labels}) of nodes and edges:

\begin{definition}[HIN]
A \emph{HIN} is a tuple $\mathcal{H} = \langle V, E, O, R, \phi, \psi \rangle$, where $V$ and $E$ are the nodes and edges of a directed multigraph, respectively; $O$ and $R$ $(|O|>1$, $|R|>1)$ are the sets of distinct node and edge types, respectively, while $\phi:V \rightarrow O$ and $\psi:E \rightarrow R$ are the mapping functions that determine the type of each node $v\in V$ and each edge $e \in E$, respectively.
\end{definition}

It is often useful to consider that each node type comes with a set of \emph{properties}, which capture useful information for the nodes of that type.\footnote{It may be useful to assign properties to edges, as well, however for notation simplicity in this work we do not consider edge properties.} Given a node~$v \in V$, we use the notation~$v.prop$ to refer to the value of the property~$prop$ on~$v$. All distinct node and edge types, along with the respective properties, can be easily derived by the HIN data and we often refer to them as the HIN's \emph{schema}.
Finally, it is often convenient (for various types of analysis) to represent the edges of a particular type $r=$ \.{o}\"{o}, where \.{o},\"{o} $\in O$ and $r \in R$, using the respective \emph{adjacency matrix} $A_r$, i.e., a \.{o} $\times$ \"{o}~matrix for which $A_r[i,j]=1$ if the $i^{th}$ node of type \.{o} is connected with the $j^{th}$ node of type \"{o}, and $A_r[i,j]=0$ otherwise.

A \emph{metapath} is a sequence of node types connected by appropriate edge types, hence corresponds to a path on a HIN schema. Given a metapath~$m$, every path in the HIN that complies to the sequence of node and edge types specified by~$m$ is an \emph{instance} of~$m$. Formally:

\begin{definition}[Metapath \& Metapath instance]
Given a HIN $\mathcal{H} = \langle V, E, O, R, \phi, \psi \rangle$, a \emph{metapath}~$m$ on~$\mathcal{H}$ is a sequence~$m = \langle o_1 \xrightarrow{r_1} o_2 \dots \xrightarrow{r_{n-1}} o_n \rangle$, where~$o_i \in O$, $r_j \in R~\forall i, j$, and~$n$ is the metapath \emph{length}. An \emph{instance} of~$m$ is any path $\langle v_1 \xrightarrow{e_1} v_2 \dots \xrightarrow{e_{n-1}} v_n \rangle$ for which $v_i \in V$, $e_j \in E$, $\phi(v_i)=o_i$, and $\psi(e_j)=r_j$ $\forall i,j$.
\end{definition}

By a common simplification~\cite{hrank, sun11}, we denote a metapath as $m = \langle o_1 o_2 \dots o_n \rangle$, without edge types, when there is only a single edge type between any pair of node types. It should be highlighted that this convention is followed here solely for the sake of simplicity; all approaches can easily accommodate edges of different types among the same pairs of nodes (e.g., since they employ separate adjacency matrices for distinct edge types).
Figure~\ref{fig:hin-example} shows an example HIN comprising $14$~nodes of $4$ types ($A$, $P$, $V$ and $T$) and $3$ edge types ($AP/PA$, $PV/VP$, and $PT/TP$). A simplified metapath of length~$3$ is~$m = \langle A P V \rangle$, and an instance of~$m$ is the path $\langle$`$Y. Vuvuli$'~`$P3$'~`$VLDB$'$\rangle$.

We also consider metapaths enhanced with constraints on node type properties, which restrict the nodes that may participate in a metapath instance. We refer to such metapaths as \emph{constrained metapaths}. The set of instances of a constrained metapath is a subset of those of the corresponding unconstrained metapath. Formally:

\begin{definition}[Constrained metapath]
\label{def:constr}
A \emph{constrained metapath} is a pair $m' = (m, C)$ where $m = \langle o_1 o_2 \dots o_n \rangle$ is a metapath and $C=\{c_1, \dots, c_n\}$ is a set of constraints on the nodes in $m$. 
\end{definition}

\begin{figure}[!t]
\vspace{-1mm}
\centering
\includegraphics[width=0.98\linewidth]{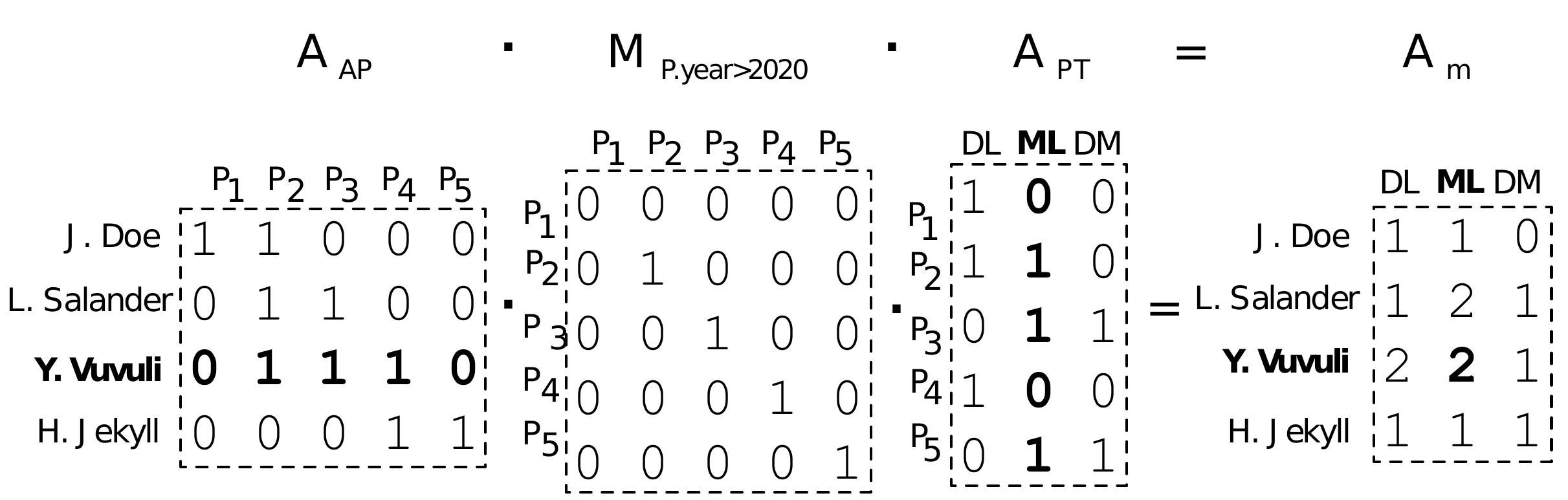}
\vspace{-2mm}
\caption{Evaluation of the $m = (\langle A P T \rangle, \{P.year>2020\})$ query on the HIN of Figure~\ref{fig:hin-example}.}\label{fig:hin-example-tables}
\vspace{-2mm}
\end{figure}

Recall the metapath $\langle A P T P A \rangle$ of the HIN in Figure~\ref{fig:hin-example}, which connects two authors that have published a paper on the same topic. A data scientist working with this HIN may need to examine only recent papers, e.g., papers published after $2000$, may use the constrained metapath $m' = (\langle A P T P A \rangle, \{P.year>2000\})$.

We now define the \emph{metapath query evaluation} (MQE) problem.

\begin{problem}[MQE]
Given a HIN $\mathcal{H} = \langle V, E, O, R, \phi, \psi \rangle$ and a (constrained) metapath $m = (\langle o_{1} o_{2} \ldots o_{n} \rangle, C)$, MQE retrieves all node pairs $\langle v_o, v_d \rangle$, such that $v_o \in o_1$, $v_d \in o_n$, and $v_o$ is connected to $v_d$ based on at least one instance of $m$. We refer to each such pair, along with the respective number of instances, as an \emph{MQE result} for~$m$. 
\end{problem}

A conventional approach to evaluate a metapath query $m$ is to calculate the matrix product~$A_m = A_{o_1o_2} A_{o_2o_3} ... A_{o_{n-1}o_n}$, where~$A_{o_io_j}$ is the adjacency matrix for the relation~$r=o_io_j$. Each non-zero element in~$A_m$ is a response to the corresponding metapath query. Given a constraint~$c_i \in C$ on object~$o_i$ in relation~$r = o_i o_j$, we apply~$c_i$ in the matrix product via the \emph{constrained} adjacency matrix~$A^{c_i}_{o_i o_j} = M_{c_i} \cdot A_{o_i,o_j}$, where~$M_{c_i}$ is a~$|o_i| \times \ |o_i|$ diagonal matrix that has element~$1$ in the diagonal only for nodes~$o_i$ satisfying~$c_i$. Figure~\ref{fig:hin-example-tables} shows an example for the metapath query~$m = (\langle A P T \rangle, \{P.year>2020\})$. Based on the MQE problem, we define the \emph{metapath query workload evaluation (MQWE)} problem:

\begin{problem}[MQWE] % we did not have time minimization above, so not here either
Given a HIN $\mathcal{H} = \langle V, E, O, R, \phi, \psi \rangle$ and a metapath query workload $W = \{m_1, \dots, m_q\}$,
MQWE is the task of efficiently computing all~$MQE$s for the metapaths in~$W$.
\end{problem}

We focus on MQWE and elaborate our approach in the following.

\section{The \thiswork{} approach}\label{sec:approach}

\thiswork, our approach for efficiently evaluating metapath query workloads, (a)~translates metapath query evaluation into a sequence of multiplications between the adjacency matrices of the edge types contributing to the respective metapaths, and (b)~applies dynamic programming to identify an efficient order of multiplications (Section~\ref{sec:planning}). Going beyond the state of the art~\cite{hrank}, \thiswork{} takes into account the inherent sparsity of the adjacency matrices to exploit fast sparse matrix multiplication algorithms, and identifies improved execution plans for each metapath using matrix multiplication cost models tailored for sparse matrices (Section~\ref{sec:sparse}). Lastly, \thiswork{} identifies sub-metapaths that reoccur due to metapath overlaps in a workload and caches the corresponding matrix multiplication results to avoid repeating them. To do so, it uses a specialized data structure, the \emph{Overlap Tree}, which keeps track of metapath overlaps, occurrence frequencies, and cache entries (Section~\ref{sec:ovtree}); \thiswork's cache management policy exploits the Overlap Tree to avoid undesirable cache replacements (Section~\ref{sec:cache-manager}).

\subsection{Efficient Multiplication Plan Selection}\label{sec:planning}

Given a metapath query, \thiswork{} generates a set of alternative execution plans, each comprising a different sequence of multiplications between the adjacency matrices of the edge types involved, as in~\cite{hrank}. Due to associativity, there are many such orderings that produce the same result, and their computational costs may differ. For example, consider the query $m=\{ \langle APT \rangle,  P.year>2020\}$ in Figure~\ref{fig:hin-example-tables}. The result $A_m=A_{AP} \cdot M_{P.year>2020} \cdot A_{PT}$ can be calculated as~$(A_{AP} \cdot M_{P.year>2020}) \cdot A_{PT}$ or~$A_{AP} \cdot (M_{P.year>2020} \cdot A_{PT})$. Since $A_{AP}$, $M_{P.year>2020}$, and $A_{PT}$ are of size $[4 \times 5]$, $[5 \times 5]$, and $[5 \times 3]$, respectively, and a standard multiplication of two matrices of size $m \times n$ and $n \times l$ requires $m \cdot n \cdot l$ operations, the first option yields~$160$ operations, while the second~$135$, which is preferable.

By this standard implementation of matrix multiplication, the cost~$C_{A_{i...j}}$ of the optimal plan for a multiplication series $A_{i...j} = A_i \cdot ... \cdot A_j$, $1 \leq i < j$ via dynamic programming (DP)~\cite{DynP}, is:

\begin{equation}
\label{eq:dynamic_programming}
  C_{A_{i...j}} = \min\limits_{i \leq k < j} \{ C_{A_{i...k}} + C_{A_{k+1...j}} + c_{A_{i...k} \cdot A_{k+1...j}} \}
\end{equation}

\noindent where~$C$ is a~$p \times p$ matrix of optimal costs and~$c_{A_{i...k} \cdot A_{k+1...j}}$ the cost of multiplying the results of sub-series~$A_{i...k}$ and~$A_{k+1...j}$, estimated by matrix dimensions in the case of standard matrix multiplication.

\subsection{Sparse Matrix Representation}\label{sec:sparse}

The technique of Section~\ref{sec:planning} has been used to identify the optimal  matrix multiplication plan for single metapath query evaluation~\cite{hrank}. However, most adjacency matrices in HINs are sparse (i.e., have few non-zero elements). Using a general-purpose matrix multiplication method on sparse matrices is inefficient. This sparsity is amplified if there are constraints in the metapath queries, which are represented by extra diagonal matrices with a small number of non-zero elements in the diagonal, whose multiplication with other matrices yields an even sparser result. Unfortunately, previous work opted for dense matrix representations~\cite{hrank}.

By contrast, we use sparse matrix structures to represent adjacency matrices and multiply such matrices taking advantage of these structures. Still, the cost to multiply sparse matrices cannot be adequately estimated from their dimensions, as it also depends on the number of non-zero elements in input and result matrices, implementation details, and memory accesses. An approximation model~\cite{SpMacho} estimates the cost of multiplying sparse matrices $X~[m \times n]$ and $Y~[n \times l]$ to yield $Z = X \cdot Y$ as:

\begin{equation}\label{eq:sparse_cost}
\hat{c_{X \cdot Y}} \approx \alpha \cdot \underbrace{(m \cdot n \cdot \rho_X)}_\text{$nonZero(X)$} + \beta \cdot \underbrace{(m \cdot n \cdot \rho_X \cdot l \cdot \rho_Y)}_\text{$\hat{N_{op}}$} + \gamma \cdot \underbrace{(m \cdot l \cdot \hat{\rho_Z})}_\text{$\hat{nonZero(Z)}$}
\end{equation}

\noindent where $\rho_X$ and $\rho_Y$ are the densities of $X$ and $Y$, $nonZero(X)$ the number of non-zero elements in $X$, $\hat{N_{op}}$ the estimated number of operations, $\hat{\rho_{Z}}$ the estimated density of $Z$, and $\hat{nonZero(Z)}$ the estimated number of non-zero elements in $Z$. Coefficients $\alpha, \beta, \gamma$ are determined by multilinear regression using least squares fit.

\begin{figure}[t]
\centering
\includegraphics[width=0.58\columnwidth]{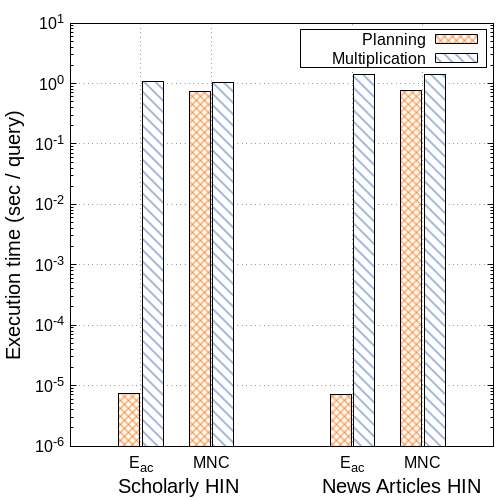}%
\caption{MNC vs the average-case density estimator $E_{ac}$ (log scale on y-axis).}
\label{fig:estimators}
\end{figure}

Assuming that the non-zero elements in $X$ and $Y$ are uniformly distributed, we compute $\hat{\rho_{Z}} = 1 - (1 - \rho_X \cdot \rho_Y)^n$ as indicated by the average-case density estimator $E_{ac}$~\cite{SpMacho, MNC}. There exist more accurate estimation algorithms~\cite{SpMacho, MNC, CohenEst, MatFast}, taking into consideration the pattern of non-zero elements in~$X$ and~$Y$, yet they incur a higher computational cost. 
We found experimentally that cost estimation techniques do not provide significant benefits in our application scenario. Figure~\ref{fig:estimators} illustrates the execution time breakdown of $E_{ac}$ against the MNC estimator~\cite{MNC}, which builds sketches to estimate the sparsity of the result. This experiment considers $2\,000$ randomly selected metapath queries with their length varying from~$3$ to~$5$ for the Scholarly and News article HINs (see Sections~\ref{sec:setup-datasets} and~\ref{sec:workloads}). Both estimators achieve very comparable results for the actual matrix multiplication execution, with $E_{ac}$ requiring a negligible amount of time for planning compared to MNC. Interestingly, both estimators produce the same multiplication plans for over $70\%$ of the examined queries. As we consider constrained metapath queries, very sparse matrices arise in the multiplication, and the plans produced by $E_{ac}$ are on par with those of more sophisticated solutions without the added cost of planning. Since the added cost of more accurate estimators does not pay off in terms of cost savings for the query types we examine, we settled for the former best-effort estimation.

In a nutshell, our sparsity-aware method produces a multiplication plan for a metapath query by the dynamic programming approach of Equation~\ref{eq:dynamic_programming}, while estimating sparse-matrix multiplication costs by Equation~\ref{eq:sparse_cost}. We use an additional~$p \times p$ matrix~$D$ to hold the densities~$\hat{\rho_Z}$ of intermediate results needed to compute this cost in each step. Besides, since \thiswork~caches intermediate matrix multiplication results (Section~\ref{sec:cache-manager}), we substitute these estimates with the (negligible) cost of retrieving a result matrix from the cache, or the cost of recomputing a result matrix if that had been previously cached and then evicted from the cache (Section~\ref{sec:ovtree}).

\subsection{The Overlap Tree}\label{sec:ovtree}

The hitherto discussed techniques focus on a single metapath query. We further enhance the evaluation of a metapath query workload by avoiding redundant computations that arise due to query overlaps. To do so, we introduce the \emph{Overlap Tree}, a dynamic data structure that organizes information about frequent metapath overlaps in the workload and their dependencies. We use this data structure, along with a cache memory that stores matrix multiplication results for the indexed overlaps, to further reduce the workload evaluation time. We provide details on cache management in Section~\ref{sec:cache-manager}.

\subsubsection{The structure}

The Overlap Tree is a dynamic data structure that keeps track of overlaps and dependencies between the already evaluated and new queries during the evaluation of a metapath query workload. It is progressively constructed by parsing query metapaths, accompanied by a cache memory that stores matrix multiplication results. At any time, it encodes the overlaps of hitherto evaluated queries, their constraints, their occurrence frequencies, and pointers to any corresponding cache entries. We first describe the structure of the Overlap Tree ignoring query constraints; we explain how the tree considers constraints in Section~\ref{sec:constraints}.

Each internal node of the tree represents an observed metapath overlap and contains the following information: 
\begin{itemize}
    \item \emph{Frequency ($f$)}: The number of occurrences of the respective overlap in the query workload until the given time point.  
    \item \emph{Cache entry pointer ($p$)}: A pointer to the cache entry containing the result of the multiplication represented by the overlap. If this result is not in the cache (i.e., has never been cached or has been cached and evicted), the pointer is null. 
    \item \emph{Multiplication cost ($c$)}: The time required to perform the matrix multiplication that corresponds to the overlap.  
    \item \emph{Matrix sparsity ($\rho$)}: The number of non-zero elements in the multiplication result matrix for that overlap.
\end{itemize}

In addition, the Overlap Tree contains one leaf node for each suffix of the metapaths of already evaluated queries (considering the metapath as a string\footnote{A query-specific character is added at the end of each metapath string to guarantee the correspondence of leaves to query suffixes in case of common prefixes. For simplicity, in all the examples we ignore this special character.}); the internal structure of each leaf is identical to that of internal nodes.\footnote{Note that $f=1$ always for leaves.} Each edge of the tree is labelled with a non-empty string, such that the concatenation of edge labels on a tree traversal from the root to a node gives the string that encodes the respective metapath overlap (or suffix in the case of leaves). 

The Overlap Tree is inspired from the \emph{generalised suffix tree}~\cite{generalised_suffix_tree_1, generalised_suffix_tree_2}, which captures the substrings of a given collection of strings; it adds pointers to cached items and information on overlap frequency, query constraints, and matrix multiplication costs. We emphasize that, as its name suggests, the Overlap tree produces a cache entry, and, for that matter, a node, only if an overlap is detected.

\begin{figure}[!t]
\centering
\includegraphics[width=\linewidth]{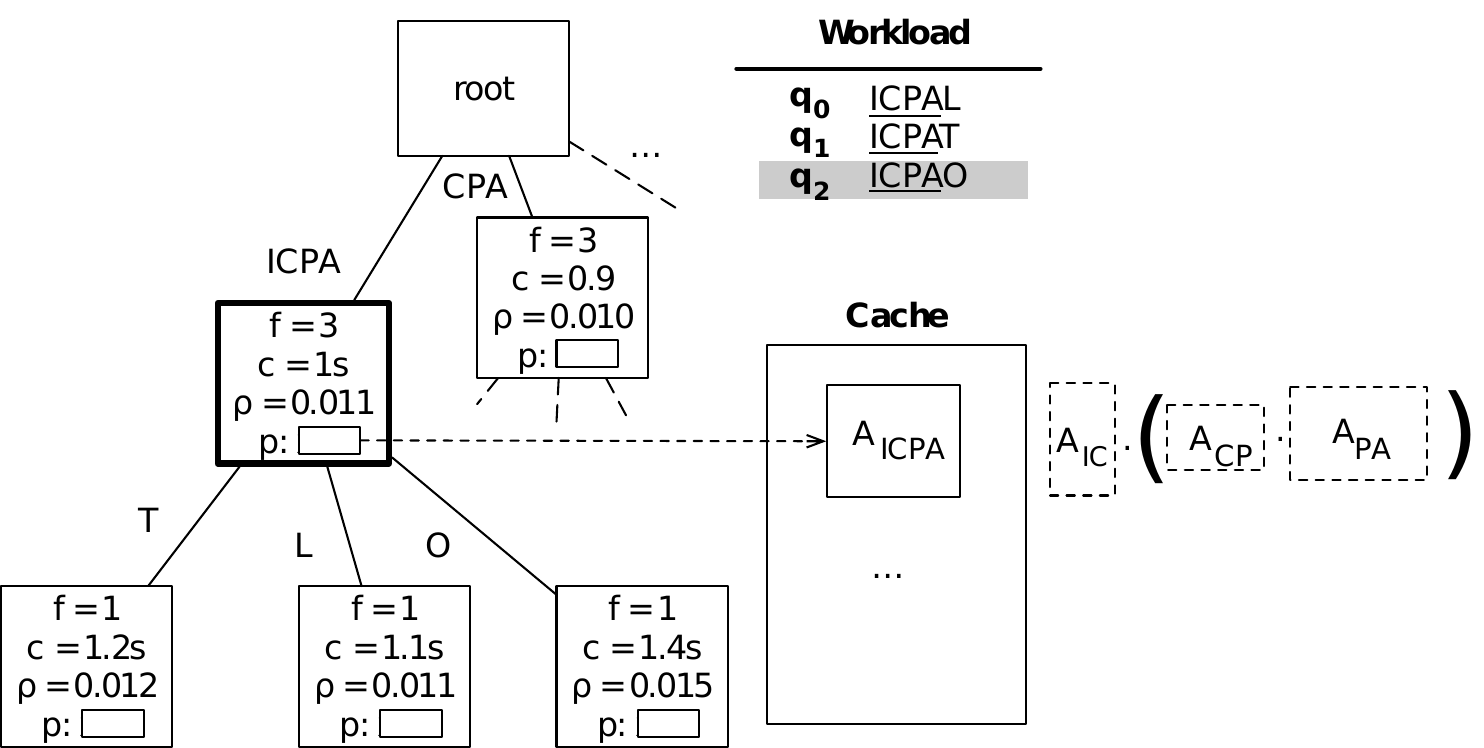}
\caption{Example workload and Overlap Tree.}\label{fig:suffix-tree}
\end{figure}

Figure~\ref{fig:suffix-tree} depicts an Overlap Tree for a workload of three metapath queries ($q_0$, $q_1$, and $q_2)$, defined on the \emph{News Articles} HIN (used in Section~\ref{sec:evaluation}).
The highlighted node represents the overlap sub-metapath~$ICPA$, occurring in all three queries, points to a cache entry for the multiplication result $A_{IC} \cdot (A_{CP} \cdot A_{PA})$, and stores the sparsity~$s$ of this cached result and the cost~$c$ to calculate it.

\subsubsection{Update}\label{sec:crup}

We construct an Overlap Tree from a metapath query workload $W = \{m_1, m_2, ..., m_q\}$ by sequentially inserting into the tree all metapath queries $m_i \in W$. For each~$m_i$ of length~$l_i=|m_i|$, we\footnote{This process is described due to its simplicity and it is not the most efficient approach to build an Overlap Tree. Approaches based on Ukkonen's algorithm~\cite{ukkonen1995} are significantly faster, yet their description is beyond the scope of this work.} insert the entire~$m_i$ and all its suffixes~ $s_i=m_i[k:l_i]$ with $0 < k < l_i$. Starting from the root, we find the longest path that matches a prefix of~$s_i$; the match ends either at an existing node or in the middle of an edge. In both cases, we let an internal node represent the detected overlap after the last matching character at position~$c$, corresponding to sub-metapath~$m_i[k:c]$; if the match ends on a leaf node, that node becomes internal, assigned dummy leaves if necessary. In case that occurs on an edge, we break that edge in two edges labelled with~$m_i[k:c]$ and~$m_i[c+1:l_i]$, and create a new internal node. Lastly, we create a new leaf node that corresponds to~$s_i$ and is connected to the last traversed internal node with a new edge. In this procedure, node frequencies are updated so as to reflect sub-metapath occurrences.

After this traversal, the cache pointers of the involved nodes may be updated. In particular, for the node representing the whole query~$m_i$, the update algorithm will attempt to cache the respective multiplication result, if the node's cache item pointer is currently null; cache insertion depends on the \emph{cache replacement policy} (Section~\ref{sec:cache-policy}). Similarly, a cache insertion will be attempted for any traversed internal node in case the respective intermediate matrix multiplication result is calculated according to the selected multiplication plan (see Sections~\ref{sec:planning}~\&~\ref{sec:sparse}). Yet even if the corresponding intermediate result is produced, it is not guaranteed to be cached, as that depends on the \emph{cache insertion policy} (Section~\ref{sec:reduce-cache}).

\subsubsection{Space Complexity} \label{sec:space}

{Consider a metapath query workload $W = \{m_1, m_2, ..., m_q\}$ and the respective Overlap Tree~$T_W$. The exact number of leaves in~$T_W$ is~$\lambda = |m_1| + |m_2| + ... + |m_q|$, since each $m_i \in W$ has exactly~$|m_i|$ suffixes.
As an internal tree node is created when and only when detecting an overlap among queries, each internal node has at least two children; in the worst case, all internal nodes have exactly two children, hence the Overlap Tree is a full binary tree having~$2\lambda-1 = O(\lambda)$ nodes.}

{Based on the internal structure of the Overlap Tree nodes, all leaves have the same memory footprint: the aggregated size of integer~$f$, two floating point numbers, $c$ and~$\rho$, and one pointer~$p$. All internal nodes, on the other hand, have the same memory footprint plus the size of the pointers to their children. Since in the worst case each internal node has two such pointers, the respective worst-case size can be easily calculated. Finally, the root has a memory footprint equal to that of two pointers, since none of the four node variables ($f,c,\rho,p$) are included in this node.}

{It is worth to highlight that for most practical scenarios, like those examined in our experiments, the size of the Overlap Tree is negligible compared to the size of the available main memory of a contemporary server. This means that the structure allows for a large size of cache to be determined, benefiting the performance of the workload evaluation process.}

\subsubsection{Handling constrained queries}\label{sec:constraints}

We have described Overlap Tree operations without considering query 
constraints (see Definition~\ref{def:constr} for details on constrained metapaths). To support constraints, we modify the internal structure of the tree nodes enriching them 
with a key-value data structure, called the \emph{constraints index}, that holds a given constraint in the form of a string (e.g., $P.year>2000$) as key and the respective variables ($f$, $p$, $c$, and $\rho$) as a composite value. The constraint index of each node can be implemented as a suitable data structure, such as a hash table index. We adapt the respective operations accordingly to accommodate this change.

Specifically, upon probing the Overlap Tree for a constrained pattern, we traverse the tree starting from the root following the exact same process described in Section~\ref{sec:crup} until we reach the ending node. Then, we probe the constraints index of the node to retrieve the respective entry (or to create it, if there is not one) and 
we update the corresponding constraints index counters accordingly; if we select to cache the result, the cache entry pointer should also be updated.

Regarding space complexity, the calculations in Section~\ref{sec:space} remain valid, yet we should also take into consideration the number of constraints each node accommodates in the constraint index. Since each leaf node represents a single, non-overlapping metapath suffix (i.e., $f=1$), each leaf stores only one key-value pair. By contrast, each internal node represents a metapath query overlap (i.e., $f>1$); in the worst case, each overlapping metapath query has a different constraint, hence the number of key-value pairs in a node equals the node's frequency~$f$.

\subsection{\thiswork-OTree Cache Management}\label{sec:cache-manager}

\begin{figure}[!t]
\centering
\includegraphics[width=0.97\linewidth]{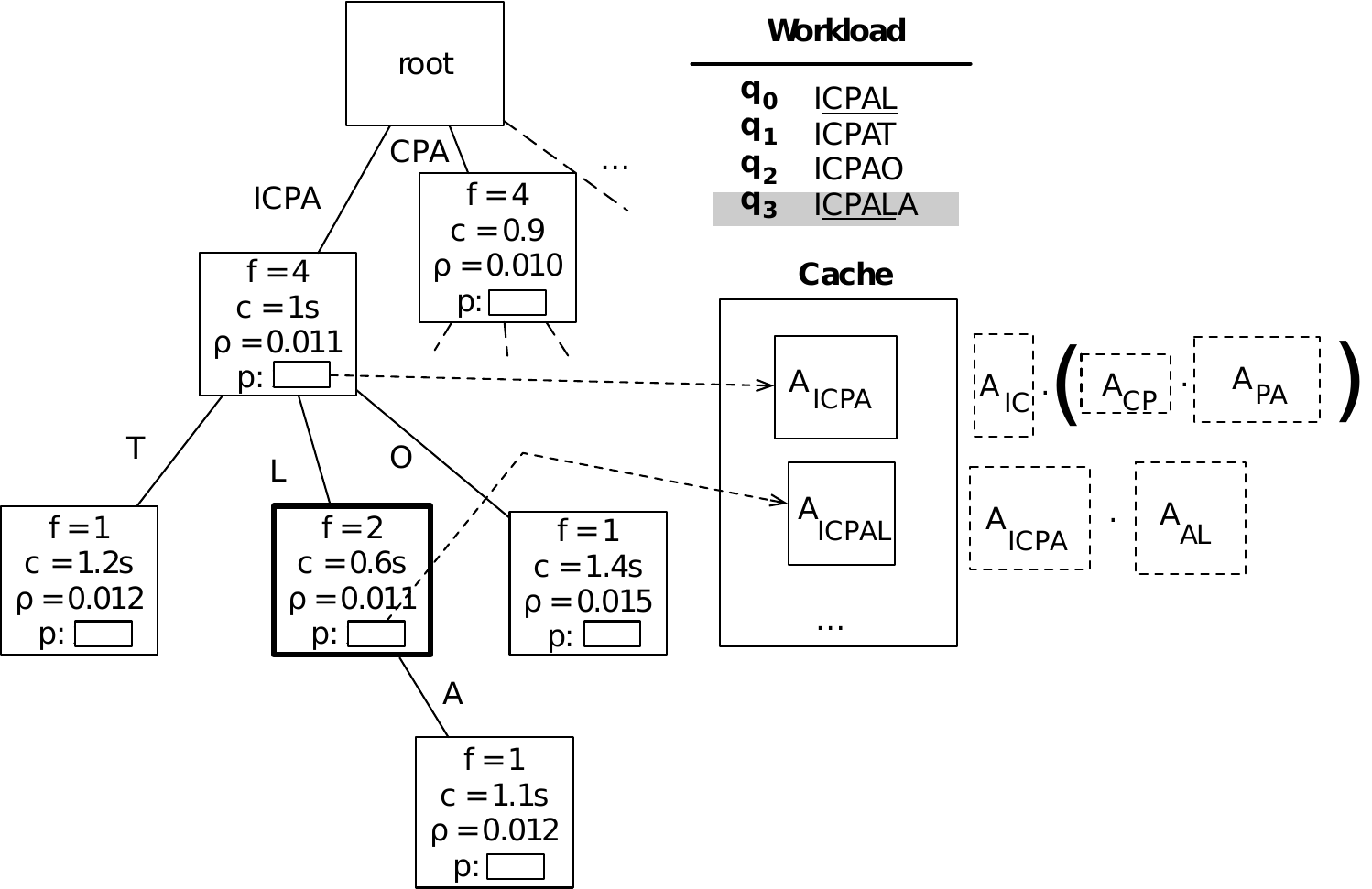}
\caption{Overlap Tree with cache entry dependencies.}\label{fig:suffix-tree-2}
\end{figure}

Evaluating a metapath query workload can benefit from caching and reusing any reappearing intermediate matrix multiplication results. However, such results are unlikely to all fit in a typical machine's main memory. We thus need a cache management policy.

We exploit the Overlap Tree to obtain a two-fold benefit: first, we use the Overlap Tree to inform a \emph{cache insertion policy favoring the insertion of frequent multiplication results}; thereby we avoid caching matrix multiplications that appear rarely, hence reduce unnecessary cache evictions (Section~\ref{sec:reduce-cache}); secondly, we use the Overlap Tree to obtain information on cached items (such as computational cost to produce, size, and dependencies) for a tailored \emph{cache replacement policy} (Section~\ref{sec:cache-policy}). In the following, we elaborate on these two uses of the Overlap Tree by \thiswork's cache management policy, the \thiswork-OTree policy.

\subsubsection{\thiswork-OTree Cache Insertion Policy}\label{sec:reduce-cache}

In any domain of knowledge, certain informative sub-metapaths are bound to occur frequently in a query workload. Such metapath repetitions may lead to redundant matrix multiplications during the evaluation of a workload. Materializing intermediate multiplication results in cache can reduce the number of redundant computations, hence improve performance. However, as there are numerous intermediate results, we need a cache insertion policy to select which ones to insert.

The Overlap Tree is helpful to that end, as each \emph{internal} node thereof encodes a sub-metapath occurring at least twice in the workload (Section~\ref{sec:crup}). Our cache insertion policy attempts to insert a cache entry for (i) the whole of~$m$ and (ii) the longest part of~$m$ that matches an internal tree node (i.e., has occurred at least twice) \emph{and} an intermediate result calculated by the matrix multiplication plan. Our cache replacement policy (Section~\ref{sec:cache-policy}) determines whether this attempt succeeds. Thereby, \thiswork reduces cache insertion attempts, focusing on frequent overlaps, which are likely to reoccur. Moreover, it selects to cache longest qualifying sub-metapath, which usually has the largest computational cost. For instance, in Figure~\ref{fig:suffix-tree-2} although~$CPA$ and~$ICPA$ are equally frequent and their intermediate results are available, only the latter is cached.

\begin{algorithm}[t]
    \SetKwInOut{Input}{inputs}
    \SetKwInOut{Output}{output}
    
    $L = 0$\\
    \ForEach{request of cache item $p$}{
        $f_p  = f_p + 1$\\
        \uIf {$p$ is in cache} {%
            $p_l = L$\\
            $h_p = f_p \cdot c_p / s_p + p_l$
        }\Else{
            \While{not enough space for $p$ in cache} {
                $L = min\{h_q | q\in cache \}$\\
                evict $q$ that satisfies $h_q = L$\\
                \ForEach{cache entry $e$ in the subtree of $q$}{
                    $c_e = c_e + c_q$\\
                    $h_e = f_e \cdot c_e / s_e + e_l$
                }
            }
            insert $p$ in cache\\
            $p_l = L$\\
            $h_p = f_p \cdot c_p / s_p + p_l$\\
            \ForEach{cache entry $e$ in the subtree of $p$}{
                $c_e = c_e - c_p$\\
                $h_e = f_e \cdot c_e / s_e + e_l$
            }
        }
    }
\caption{\thiswork-OTree cache replacement policy}
\end{algorithm}

\subsubsection{\thiswork-OTree Cache Replacement Policy}\label{sec:cache-policy}

A straightforward way to cache matrix multiplication results is to use a generic policy, like LRU~\cite{aho1971}, which discards the \emph{least recently used} cached item first. LRU works well when all items have similar (a)~cost to fetch and (b)~size. However, neither of these conditions holds for metapath query workloads, as (i)~fetching an item in the cache involves matrix multiplications, whose cost differs among items, and (ii)~item size varies depending on the result matrix density. In effect, one may adopt a size- and cost-aware cache replacement policy, as in web server caching~\cite{LUV, GDStar, Jin2000}. Popularity-aware Greedy Dual-Size (PGDS)~\cite{Jin2000} measures the utility of a cache entry $e$ as $u_{e} = f_{e} \cdot \frac{c_{e}}{s_{e}}$, where $f_e$ is the frequency of~$e$, $c_e$ its cost, and~$s_e$ its size. We may obtain~$f_e$ of each cache entry from the corresponding tree counter, $c_e$ as the estimated cost of the corresponding matrix multiplication, and~$s_e$ as its size. Let~$E$ be all cache entries. Initially, each entry~$e_i \in E$ has utility value~$h_{e_i} = u_{e_i}$. In case of cache saturation, we evict the entry with~$\min\{h_{e_i}\}$, denoted as~$H_{\min\\}$ and reduce the utility values of all other cache items by~$H_{\min\\}$. Thus, recently cached items retain a higher fraction of their initial utility and are hence less likely to be evicted. To avoid subtracting~$H_{\min\\}$ from items in the cache, PGDS uses an inflation variable~$L$: it adds~$H_{\min\\}$ to~$L$ at each eviction, and increases the utility of any cache hit by~$L$.

We employ an enhanced cache policy that builds on PGDS by considering cache entry interdependence: the cost~$c_e$ to re-calculate and re-fetch an entry~$e$ in the cache may be smaller than what PGDS estimates, because part of the calculation may be obtained by another cache entry, $e'$, as cache entries follow a hierarchy: given an internal Overlap Tree node~$n_p$ and the set of nodes in its sub-tree, $S(n_p)$, the cache entry of each~$n_i \in S(n_p)$ can exploit that of~$n_p$. Thus, each time we insert a cache entry~$p$ in an internal node, we \emph{subtract} its cost~$c_p$ from that of cached items in its sub-tree, as these are cheaper to recompute given~$p$. In reverse, when we evict an entry~$q$ from the cache, we \emph{reinstate} its cost~$c_q$ to that of items in its sub-tree. Algorithm~1 
outlines this approach.

Figure~\ref{fig:suffix-tree-2}~shows the Overlap Tree of Figure~\ref{fig:suffix-tree} after inserting a new metapath, $ICPALA$. Assuming that the pointer~$p$ in the highlighted node for~$ICPAL$ was previously null, we insert a new cache item, containing the respective multiplication result, and update~$p$ to point to it. Still, since the traversal passed through the node for~$ICPA$, we can use its cached result to calculate the result for~$ICPAL$. In case the cached item for~$ICPA$ is evicted, we reinstate its cost to that of cached items in its sub-tree, i.e., to~$ICPAL$, to reflect the cost of its calculation without using the entry for~$ICPA$.

\section{Experimental Evaluation}\label{sec:evaluation}

In this section, we present an extensive experimental evaluation of \thiswork. We describe the experimental setup in Section~\ref{sec:setup}, showcase the performance of \thiswork against state-of-the-art approaches for single metapath query evaluation in Section~\ref{sec:eval-sota}, and present experiments against two baselines that take advantage of a caching mechanism for metapath workloads in Section~\ref{sec:against_rivals}; we investigate various cache replacement policies in Section~\ref{sec:against_policies}.

\subsection{Setup}\label{sec:setup}

\subsubsection{Datasets.}\label{sec:setup-datasets}

\begin{figure}[t]
\centering
\begin{subfigure}{0.4\columnwidth}
\includegraphics[width=\columnwidth]{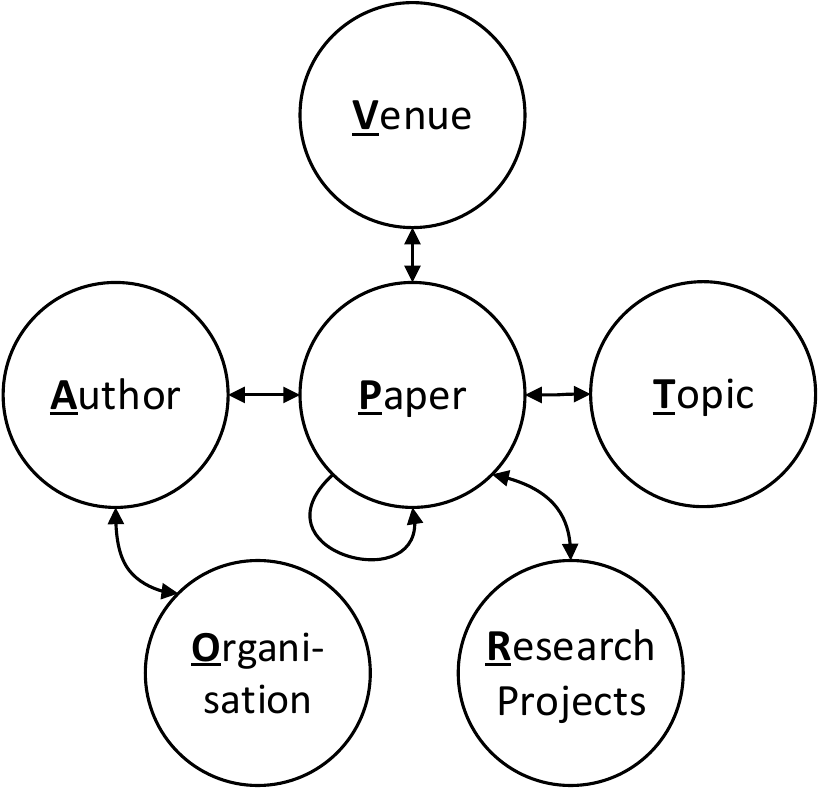}%
\caption{\dblp}%
\label{fig:dataset_schemas_DBLP}%
\end{subfigure}\hspace{0.15\columnwidth}%
\begin{subfigure}{0.4\columnwidth}
\includegraphics[width=\columnwidth]{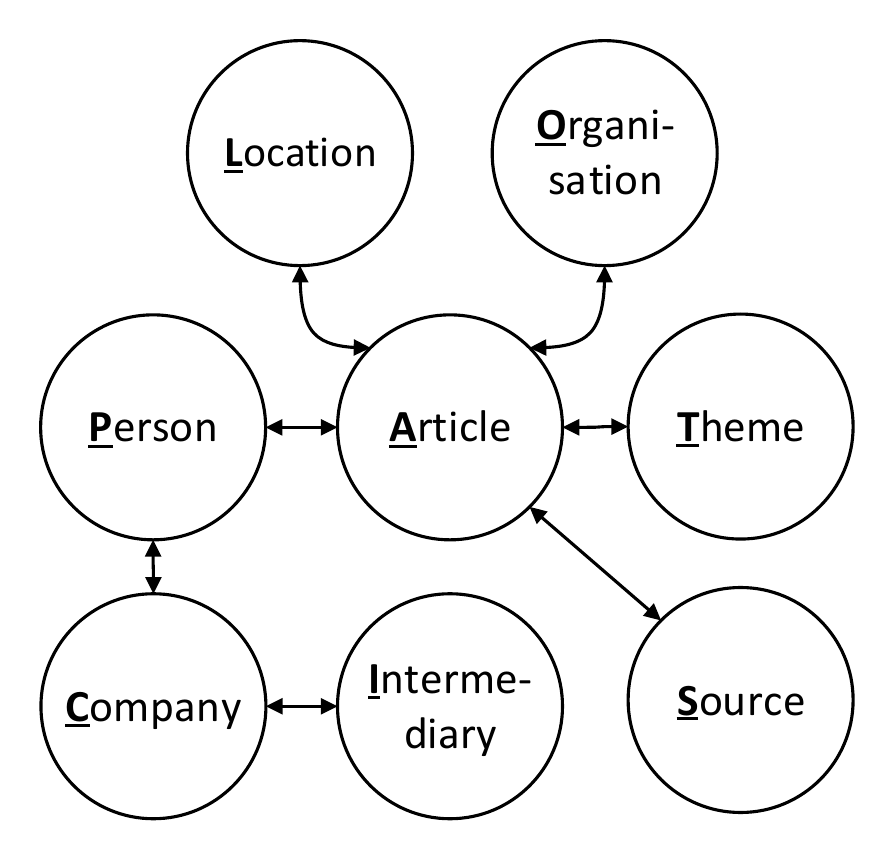}%
\caption{\gdelt}\label{fig:dataset_schemas_GDELT}%
\end{subfigure}\hfill%
\caption{Implied schemas of experimental HINs.}\label{fig:hin_schemas}
\end{figure}

We use the following data sets:
\begin{itemize}
\item \textbf{\dblp.} A dataset based on AMiner's DBLP Citation Dataset~\cite{aminer}, enriched
with projects funded by the EU under the Horizon 2020 programme, taken from Cordis\footnote{\url{https://data.europa.eu/euodp/en/data/dataset/cordisH2020projects}}, including entities for \texttt{Papers}~(\texttt{P}), \texttt{Authors}~(\texttt{A}), \texttt{Organisations}~(\texttt{O}), \texttt{Venues}~(\texttt{V}), \texttt{Topics}~(\texttt{T}), and \texttt{Research Projects}~(\texttt{R}). Figure~\ref{fig:hin_schemas}(a) shows the implied schema.
\item \textbf{\gdelt.} This is based on news articles published on April 2016 and their associated entities from the GDELT Project~\cite{Leetaru2013}, enriched with information for people involved in the Panama Papers scandal and their related companies and intermediaries from the Offshore Leaks database\footnote{\url{https://offshoreleaks.icij.org/pages/database}}). The integration was performed by linking Panama Papers officers with persons in GDELT. 
The resulting HIN comprises \texttt{Articles}~(\texttt{A}), \texttt{Organisations~(\texttt{O})}, \texttt{Persons}~(\texttt{P}), \texttt{Locations} (\texttt{L}), \texttt{Sources}~(\texttt{S}),  \texttt{Themes}~(\texttt{T}),   
\texttt{Compa\-nies}~(\texttt{C}), and their \texttt{Inter\-mediaries}~(\texttt{I}). Its implied schema is shown in~Figure~\ref{fig:hin_schemas}b.
\end{itemize}

\begin{table}[b]
\centering
\scriptsize
\caption{Details of experimental HINs.}\label{tbl:dataset_details}
\begin{tabular}{ccccc}
    \toprule
    \textbf{Dataset} & \textbf{Contained Core} & \textbf{Nodes} & \textbf{Edges} & \textbf{Avg. Degree}\\ 
    & \textbf{Entity Nodes} & \textbf{(in millions)} & \textbf{(in millions)} & \\
    \midrule
    & $60\%$ & $8.65$ & $108.43$ & $25.06$\\
    \textbf{\dblp}& $80\%$ &$10.44$ & $148.09$ & $28.36$\\
    & $100\%$ & $12.15$ & $190.99$ & $31.43$\\
    \midrule
    & $60\%$ & $7.71$ & $173.38$ & $44.92$\\
    \textbf{\gdelt} & $80\%$ & $10.06$ & $233.33$ & $46.37$\\
    & $100\%$ & $12.43$ & $298.11$ & $47.95$\\
    \bottomrule
\end{tabular}
\end{table}

To assess scalability, we consider two additional splits of different sizes for the aforementioned HINs, containing $60\%$ and $80\%$ of their core entities, i.e. papers and articles, while preserving nodes and edges connected with them. On \gdelt, we first keep across splits all articles connected with companies, as otherwise few or no companies may be present in the final dataset, and we then augment this core set to reach the number of articles needed. Table~\ref{tbl:dataset_details} shows their statistics, i.e., number of nodes and edges, 
while Table~\ref{tbl:more_dataset_details}~collects the number of nodes and edges per type for each dataset split.
Note that, in all cases, we consider bi-directional relationships between entities.
These HINs are two orders of magnitude larger than those considered in previous recent studies~(e.g. \cite{hrank}).

\begin{table}[t]
\centering
\scriptsize
\caption{Number of nodes/edges per type.}\label{tbl:more_dataset_details}
\begin{tabular}{c|cc|cc}
    \toprule
    & \textbf{Node} & \textbf{Million nodes per split} & \textbf{Edge} & \textbf{Million edges per split}\\
    & \textbf{Type} & $\mathbf{(60\%~/~80\%~/~100\%)}$ & \textbf{Type} & $\mathbf{(60\%~/~80\%~/~100\%)}$\\
    \midrule
    \parbox[t]{2mm}{\multirow{6}{*}{\rotatebox[origin=c]{90}{\textbf{\dblp}}}} 
    & \texttt{P} & $2.936/3.915/4.894$ & \texttt{PP} & $16.367/29.236/455.641$\\
    & \texttt{A} & $3.100/3.781/4.398 $ & \texttt{AP/PA} & $17.916/23.898/29.869$\\
    & \texttt{O} & $2.488/2.608/2.706 $ & \texttt{OA/AO} & $14.921/15.976/16.832$\\
    & \texttt{V} & $0.009/0.009/0.010$ & \texttt{VP/PV} & $5.247/6.994/8.743$\\
    & \texttt{T} & $0.117/0.125/0.132$ & \texttt{TP/PT} & $53.977/71.974/89.974$\\
    & \texttt{R} & $0.001/0.001/0.001$ & \texttt{RP/PR} & $0.008/0.010/0.013$\\
    \midrule
    \parbox[t]{2mm}{\multirow{3}{*}{\rotatebox[origin=c]{90}{\textbf{\gdelt}}}} & \texttt{A} &$4.394/5.859/7.324$ & \texttt{IC/CI} & $0.010/0.010/0.010$\\
    & \texttt{O} &$1.153/1.486/1.829$ & \texttt{OA/AO} & $25.786/35.186/45.584$\\
    & \texttt{P} &$1.953/2.465/2.995$ & \texttt{PA/AP} & $32.902/44.334/57.125$\\
    & \texttt{L} &$0.172/0.202/0.229$ & \texttt{LA/AL} & $31.767/42.985/55.319$\\
    & \texttt{T} &$0.014/0.016/0.17$ & \texttt{TA/AT} & $73.184/97.618/122.750$\\
    & \texttt{S} &$0.022/0.025/0.030$ & \texttt{SA/AS} & $9.714/13.185/17.306$\\
    & \texttt{C} &$0.005/0.005/0.005$ & \texttt{CP/PC} & $0.014/0.014/0.014$\\
    & \texttt{I} &$0.001/0.001/0.001$ & \\
    \bottomrule
\end{tabular}
\end{table}

\subsubsection{Query Workloads}\label{sec:workloads}

Due to the lack of available real-world workloads, we create synthetic query workloads simulating the query patterns of multiple data scientists posing queries, each exploring different aspects of a certain entity (e.g., a particular author in the Scholarly HIN) via consecutive metapath queries related to it. The entity of interest is determined by an equality constraint. We refer to such a set of constrained metapath queries as a \emph{query session}. We generate such queries by randomly choosing a metapath and a constraint applied on it. In each step, we either (a) randomly pick a new constraint and start a new session, with \emph{session restart probability}~$p$, or (b) continue using the same constraint and pick a different metapath, with probability~$1-p$. Indicatively, $p=0.1$ results in sessions each containing on average~$10$ metapath queries, which is a reasonable length. Eventually, we shuffle all queries in a workload to simulate a real-world environment. For all random selections we used a uniform or a zipfian probability distribution, the former as default. Each query workload contains~$500$ metapath queries with length ranging from~$3$ to~$5$. We repeat each experiment with~$10$ different query workloads, and report the average evaluation time. All the query workloads are openly available.\footnote{\url{https://github.com/atrapos-hin/artifacts/tree/main/workloads}}

\subsubsection{Methods}

We juxtapose the following methods:
\begin{itemize}
    \item \emph{HRank}: the approach of~\cite{hrank} in its original configuration. 
    \item \emph{Neo4j}~\cite{webber2012}: a well-established graph database. We convert each metapath query to its corresponding Cypher\footnote{Neo4j's native query language: \url{https://neo4j.com/developer/cypher-query-language/}} query, using the parameters' syntax\footnote{\url{https://neo4j.com/docs/cypher-manual/current/syntax/parameters/}} to express query constraints, enabling Neo4j to cache query execution plans.
    \item \emph{HRank Sparse (HRank-S)}: a version of HRank adjusted to exploit sparse matrix representations and the cost estimation for sparse matrix multiplications of Section~\ref{sec:sparse}.
    \item \emph{Cache-based Baseline-1 (CBS1)}: an adaptation of HRank-S using an LRU cache to store \emph{final} metapath query results.
    \item \emph{Cache-based Baseline-2 (CBS2)}: an extension of \emph{CBS1} that caches all \emph{intermediate} matrix multiplication results.
    \item \emph{\thiswork{}}: our approach that uses the Overlap Tree to detect frequent query overlaps and cache entry dependencies.
\end{itemize}

\emph{HRank} and \emph{Neo4j} are state-of-the-art approaches for single metapath query evaluation which we apply on query workloads. \emph{Neo4j} has its own internal caching mechanism. As discussed in Section~\ref{sec:eval-sota}, these approaches perform poorly even for small datasets and do not scale to large HINs. Therefore, we also use the three baselines (\emph{HRank-S}, \emph{CBS1}, and \emph{CBS2}), which take advantage of sparse matrix representations and simple caching mechanisms.
Note that, all intermediate results evicted from the cache could, in theory, be stored on disk instead of discarding them, acting like a second level cache mechanism.
However, this applies to all cache-based methods and, hence, does not impact their relative comparison.

\subsubsection{Evaluation Setting}

We implemented\footnote{Code available at \url{https://github.com/atrapos-hin/artifacts}} all methods in C++, using the Eigen Linear Algebra Library~\cite{eigenweb} for matrix multiplications. We opted for one of the most popular sparse matrix data structures, which stores non-zero values and their positions in a column-major order, the compressed sparse column layout~(CSC).\footnote{CSC is the recommended sparse matrix layout in Eigen: \url{https://eigen.tuxfamily.org/dox/group__TopicStorageOrders.html}}
All source codes were compiled with the GNU Compiler (G++ version 9.3.0)
~with o3 optimization enabled and the NDEBUG preprocessor directive that suppresses assertions to reduce execution time. 
All cache-based implementations are set to never store items that exceed a size threshold equal to~$80\%$ of the available cache size. 
This rule safeguards against evicting many useful results to store just a single large one (indicatively, the rule applied in less than~$0.1\%$ of the queries for all cache-based approaches in both datasets considering the default setup with $4$GB cache).

\begin{table}[!t]
\centering
\scriptsize
\caption{Experimental parameters.}\label{tbl:parameters}
\begin{tabular}{ccc}
    \toprule
    \textbf{Parameter} & \textbf{Range of values}\\ \midrule
    \textbf{Cache size (MB)} & $1024, 2048, \bm{4096}, 8128, 16256$\\
    \textbf{Dataset size (core entity nodes)} & $\bm{60\%}, 80\%, 100\%$\\
    \textbf{Session restart probability} & $0.04, 0.06, \bm{0.08}, 0.1, 0.12$\\
    \textbf{Workload distribution} & \textbf{Uniform}, Zipfian\\
    \bottomrule
\end{tabular}
\end{table}

All experiments ran on an AMD Ryzen Threadripper 3960X CPU machine @ 3.50GHz with $256$GB RAM on Ubuntu 20.04 LTS. 
All considered approaches were able to utilise all the available memory.
Specifically, the Neo4j approach runs on the Neo4j 4.2.5 Community Edition, configured to allocate $31$GB of JVM heap size and $205$GB of page size, as recommended.\footnote{\url{https://neo4j.com/docs/operations-manual/current/tools/neo4j-admin-memrec/}} The remaining memory can be allocated by Neo4j from the OS (native memory) hence all $256$GB of memory could be utilised in total (alternative configurations had no significant impact on performance).
In addition, \thiswork~and its competitors load data from CSV files into the main memory; 
this ingestion time is not included in the reported query execution time.
Table~\ref{tbl:parameters} lists the parameters we investigate and the range of values for each parameter, with default values in bold.

\subsection{Evaluation against single-query methods}\label{sec:eval-sota}

\begin{figure}[t]
\centering
\begin{subfigure}{.5\columnwidth}
\centering
\includegraphics[width=1\columnwidth]{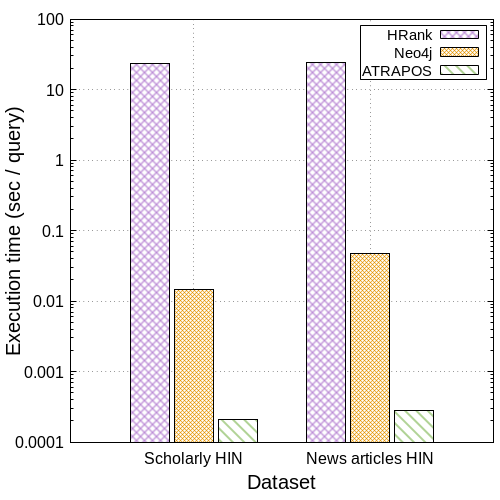}%
\caption{ Vs. HRank \& Neo4j \\(dataset subsets, log scale).}\label{fig:hrank-neo4j}%
\end{subfigure}\hfill%
\begin{subfigure}{.5\columnwidth}
\centering
\includegraphics[width=1\columnwidth]{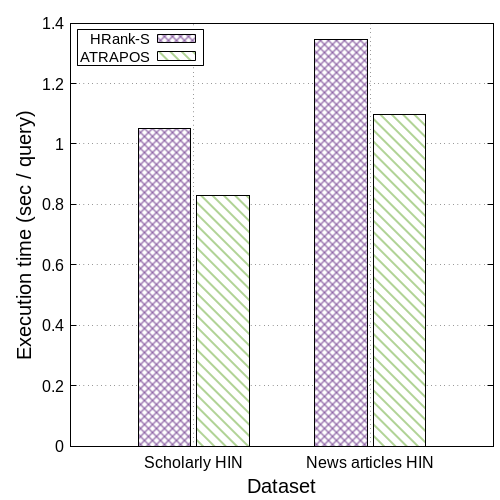}%
\caption{ Vs. HRank-S (full dataset \\sizes, linear scale)}\label{fig:hrank-s}%
\end{subfigure}\hfill%
\caption{Evaluation against single-query methods.}\label{neo4j-comparison}%
\end{figure}

First, we assess \thiswork against the dense-matrix HRank and Neo4j, using subsets of the \dblp ($37k$ nodes, $120k$ edges) and the \gdelt ($37k$ nodes, $210k$ edges), as HRank and Neo4j cannot handle the full datasets due to their memory requirements. Figure~\ref{fig:hrank-neo4j}~presents our results on execution time per query in logarithmic scale. \thiswork surpasses contestants on both datasets. Neo4j is over one order of magnitude slower than \thiswork, while HRank is over two orders of magnitude slower than Neo4j. 

Then, we examine \thiswork~against HRank-S, using full size HINs; HRank-S is a variant of HRank that takes advantage of sparse matrix representations. Both \thiswork and HRank-S use the same sparse matrix representations and the respective multiplication cost estimation described in Section~\ref{sec:sparse}. However, \thiswork~leverages metapath query overlaps using a~$4$GB cache. Figure~\ref{fig:hrank-s}~presents our findings on execution time per query. \thiswork~outperforms HRank-S in both datasets with a speedup of~$27\%$ on the \dblp and~$23\%$ in the \gdelt.

\subsection{Evaluation against baseline caching}\label{sec:against_rivals}

While no existing approach exploits a caching mechanism to accelerate the evaluation of a metapath query workload, evaluating \thiswork{} solely against non-caching approaches cannot be fair. Thus, we introduce CBS1 and CBS2, two HRank-S variants that employ basic caching mechanisms (Section~\ref{sec:setup}) and assess \thiswork{} against them. We include HRank-S in the same experiment for reference. We first study performance vs. cache size and dataset size. Then, we investigate workload parameters, i.e., the session restart probability~$p$ and the selection of metapaths and constraints.

\begin{figure}[t]
\centering
\begin{subfigure}{.5\columnwidth}
\centering
\includegraphics[width=.95\columnwidth]{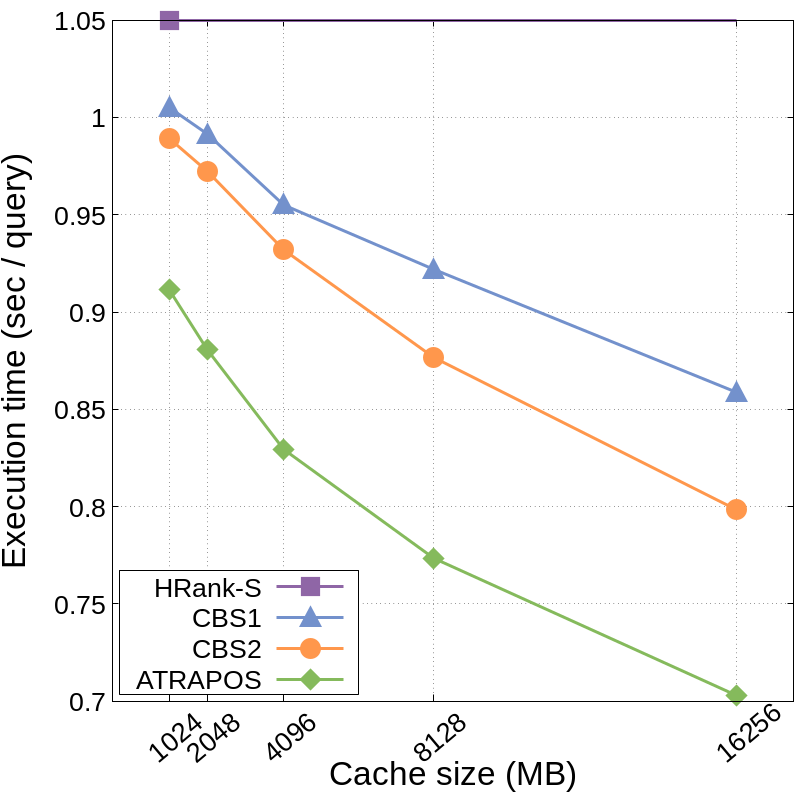}%
\caption{\dblp}\label{subfiga}%
\end{subfigure}\hfill%
\begin{subfigure}{.5\columnwidth}
\centering
\includegraphics[width=.95\columnwidth]{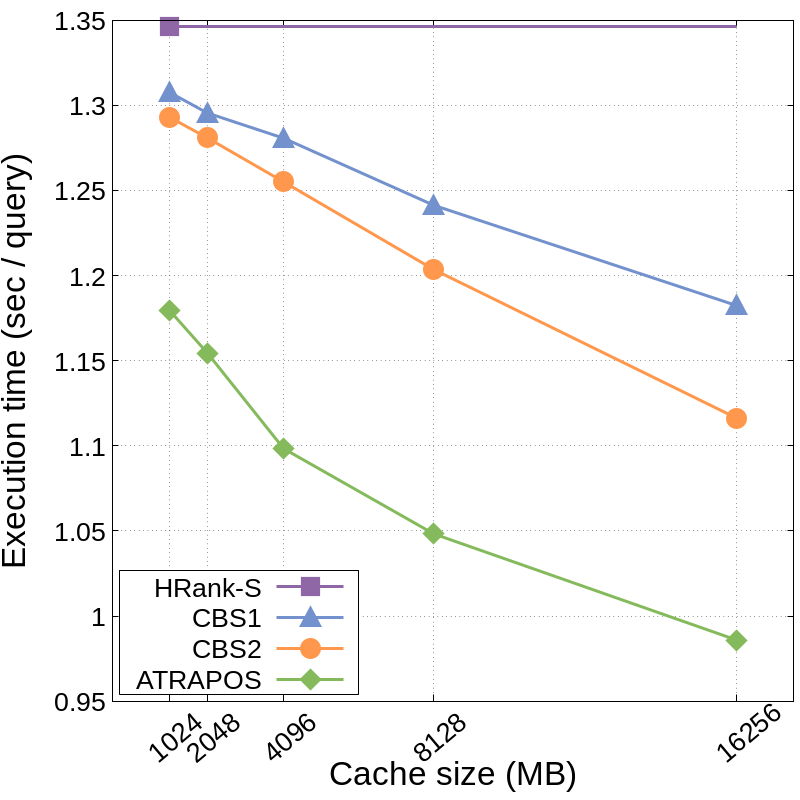}%
\caption{\gdelt}\label{subfigb}%
\end{subfigure}\hfill%
\caption{Evaluation against baseline caching approaches with varying cache size.}
\label{fig:rivals-cache}
\end{figure}

\subsubsection{Varying cache size.}\label{sec:against_rivals-cache_size}

Figure~\ref{fig:rivals-cache} presents the average execution time per query vs. cache size. As HRank-S is oblivious to caching, its execution time is independent of cache size. All cache-based approaches outperform HRank-S in both datasets, while their advantage grows with cache size. CBS1 achieves faster times than HRank-S by virtue of exploiting repetitive metapath queries; as it caches only query results, it is slower than CBS2 and \thiswork.
\thiswork{} outperforms other contestants on both datasets, as it leverages the Overlap Tree to identify frequent metapath overlaps and select matrix multiplication results for caching. Its cache management policy refrains from inserting all intermediate results into the cache, therefore its advantage over CBS2 is accentuated in small cache sizes ($2$GB and $4$GB).

\begin{figure}[t]
\begin{subfigure}{.5\columnwidth}
\centering
\includegraphics[width=.95\columnwidth]{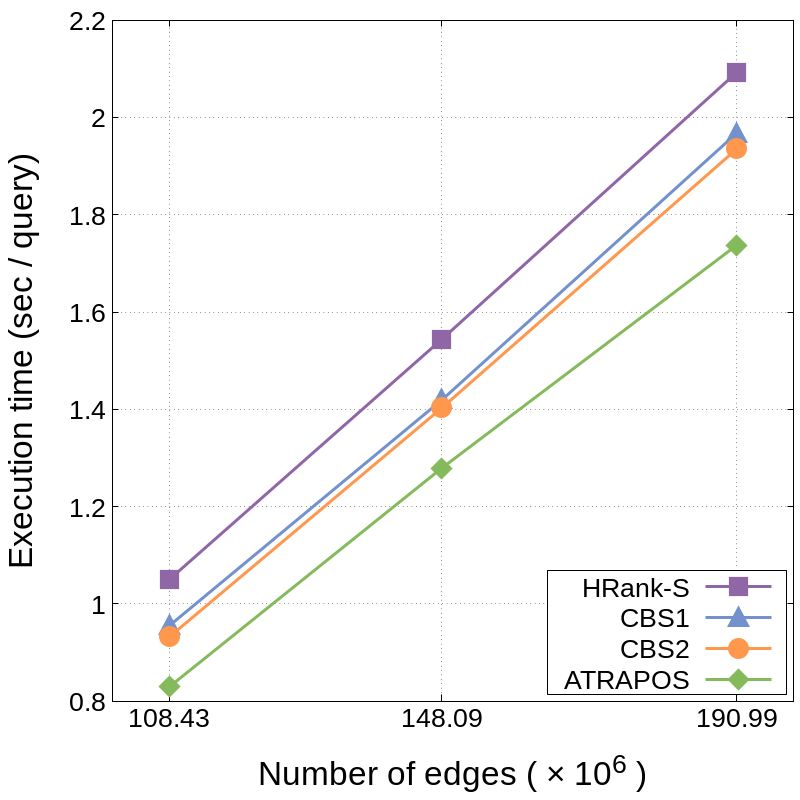}%
\caption{\dblp}\label{subfiga}%
\end{subfigure}\hfill%
\begin{subfigure}{.5\columnwidth}
\centering
\includegraphics[width=.95\columnwidth]{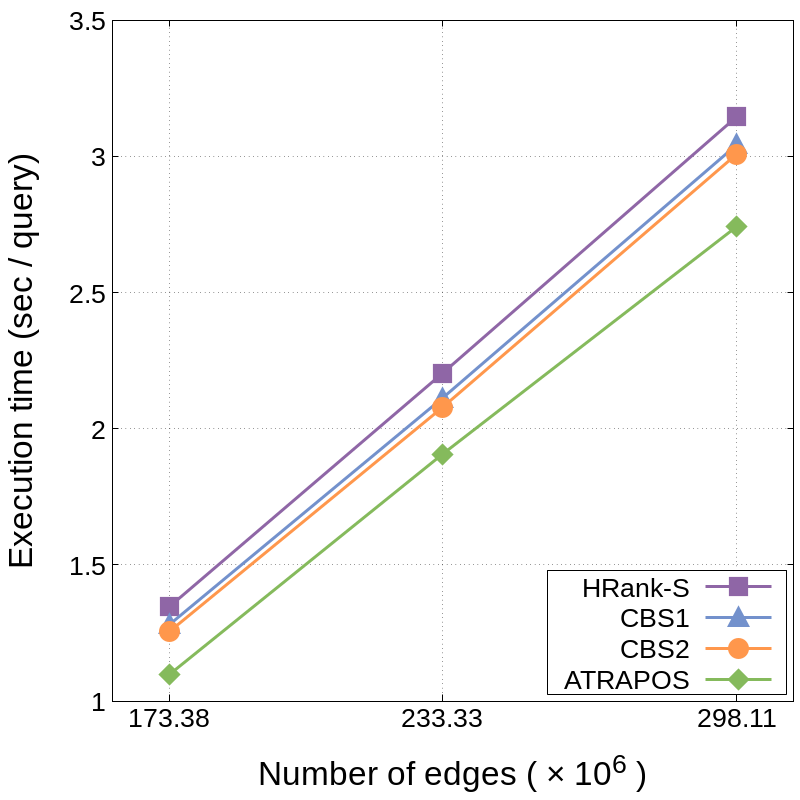}%
\caption{\gdelt}\label{subfigb}%
\end{subfigure}\hfill%
\caption{Evaluation against baseline caching approaches with varying dataset size.}
\label{fig:rivals-data-size}
\end{figure}

\subsubsection{Varying dataset size.}

Figure~\ref{fig:rivals-data-size} illustrates the average execution time per query vs. network size. The~$x$ axis contains the number of edges (in millions) for each dataset split (Section~\ref{sec:setup-datasets}). All methods scale linearly with the number of edges. All three cache-based approaches outperform HRank-S, with the difference being more prominent in the \dblp. \thiswork clearly outperforms CBS1 and CBS2 on all dataset sizes.

\begin{figure}[t]
\centering
\begin{subfigure}{.5\columnwidth}
\centering
\includegraphics[width=.95\columnwidth]{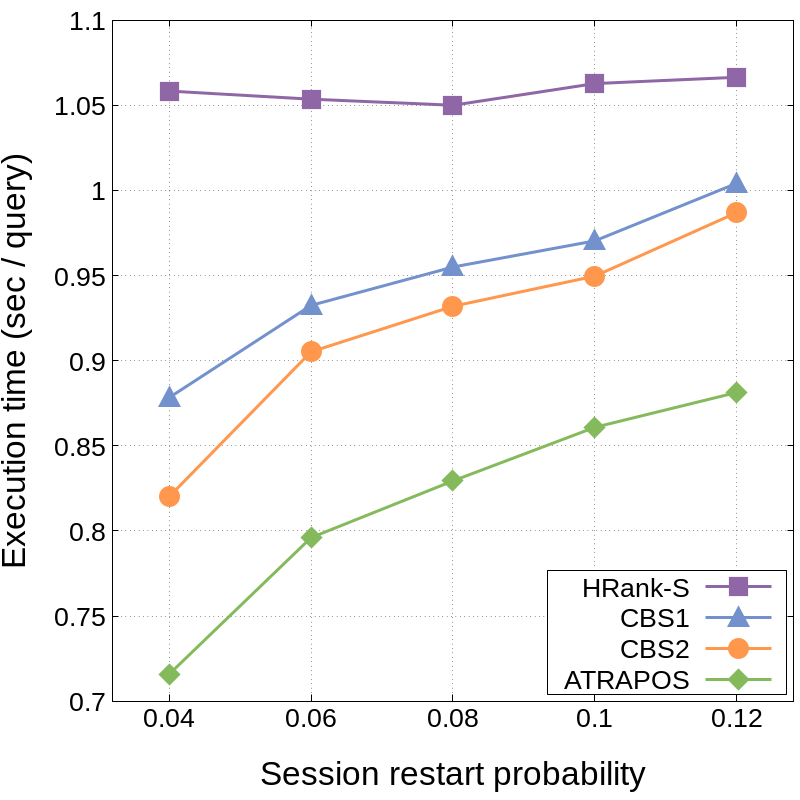}%
\caption{\dblp}\label{subfiga}%
\end{subfigure}\hfill%
\begin{subfigure}{.5\columnwidth}
\centering
\includegraphics[width=.95\columnwidth]{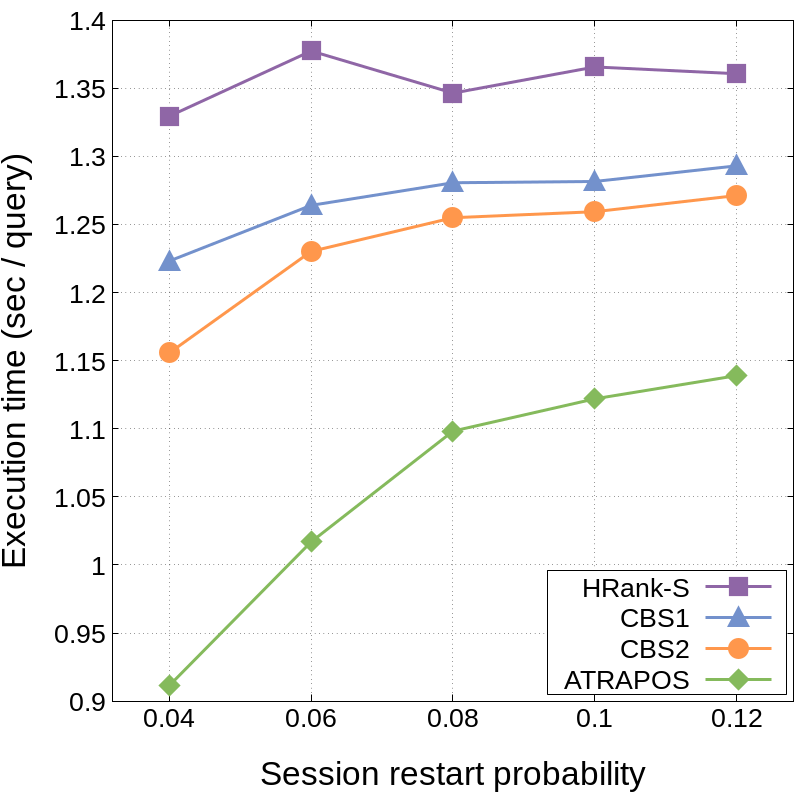}%
\caption{\gdelt}\label{subfigb}%
\end{subfigure}\hfill%
\caption{Evaluation against baseline caching approaches varying session restart probability.}\label{fig:rivals-sessions}
\end{figure}

\subsubsection{Varying session restart probability $p$.}\label{sec:against_rivals-sessions}

Figure~\ref{fig:rivals-sessions} shows the improvement in execution time over HRank-S for CBS1, CBS2 and \thiswork{} as the session restart probability~$p$ falls. As~$p$ grows, the performance of all cache-based methods degrades, as there are fewer queries per session on average, hence fewer overlaps to exploit. Overall, \thiswork{} outperforms its competitors. Among the rival methods, CBS2 surpasses CBS1, as it exploits intermediate results.

\begin{figure}[t]
\begin{subfigure}{.5\columnwidth}
\centering
\includegraphics[width=.95\columnwidth]{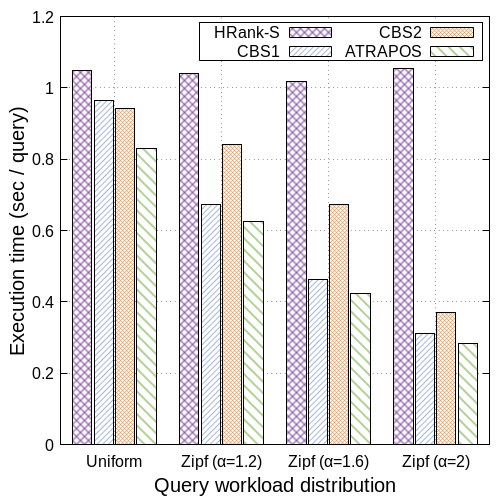}%
\caption{\dblp}\label{subfiga}%
\end{subfigure}\hfill%
\begin{subfigure}{.5\columnwidth}
\centering
\includegraphics[width=.95\columnwidth]{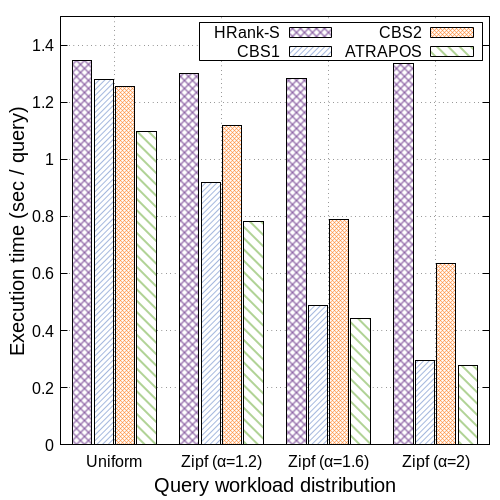}%
\caption{\gdelt}\label{subfigb}%
\end{subfigure}\hfill%
\caption{Evaluation against baseline caching approaches with the queries following the Zipf distribution.}\label{fig:rivals-zipf}
\end{figure}

\subsubsection{Zipfian query workloads.}

Figure~\ref{fig:rivals-zipf} shows execution times per query as we vary the distribution from which we select metapaths and constraints. Apart from the uniform distribution, used in other experiments, we consider Zipfian distributions varying the scaling parameter~$\alpha$. HRank-S achieves comparable execution times in all settings as it does not take advantage of repetitions, while cache-based approaches improve as~$\alpha$ grows. As~$\alpha$ rises, the probability of queries being repeated also increases, while smaller values indicate a heavier tail in the probability distribution. \thiswork outperforms competitors in all scenarios: it outperforms CBS1 by exploiting query overlaps on top of the result that CBS1 caches. The difference is more notable with~$\alpha = 1.2$ and~$\alpha = 1.6$, yet \thiswork surpasses CBS1 even with~$\alpha = 2$, whereby a few queries are repeated in the workload. CBS2 exhibits significantly larger execution times with Zipfian distributions, as caching intermediate results bars it from exploiting all query repetitions. 

\begin{figure}[t]
\begin{subfigure}{.5\columnwidth}
\centering
\includegraphics[width=.95\columnwidth]{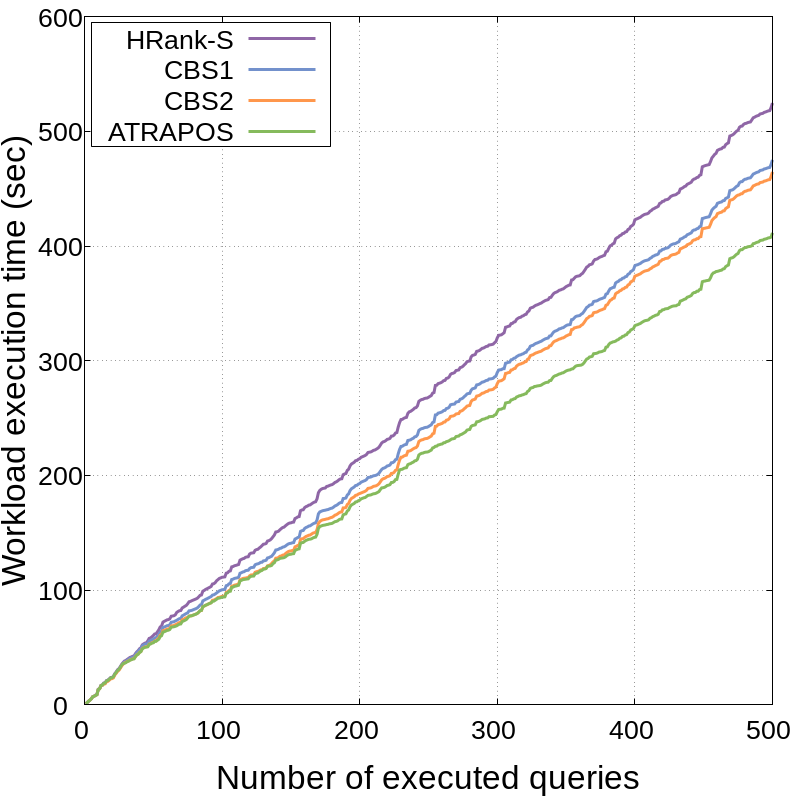}%
\caption{\dblp}%
\label{subfiga}%
\end{subfigure}\hfill%
\begin{subfigure}{.5\columnwidth}
\centering
\includegraphics[width=.95\columnwidth]{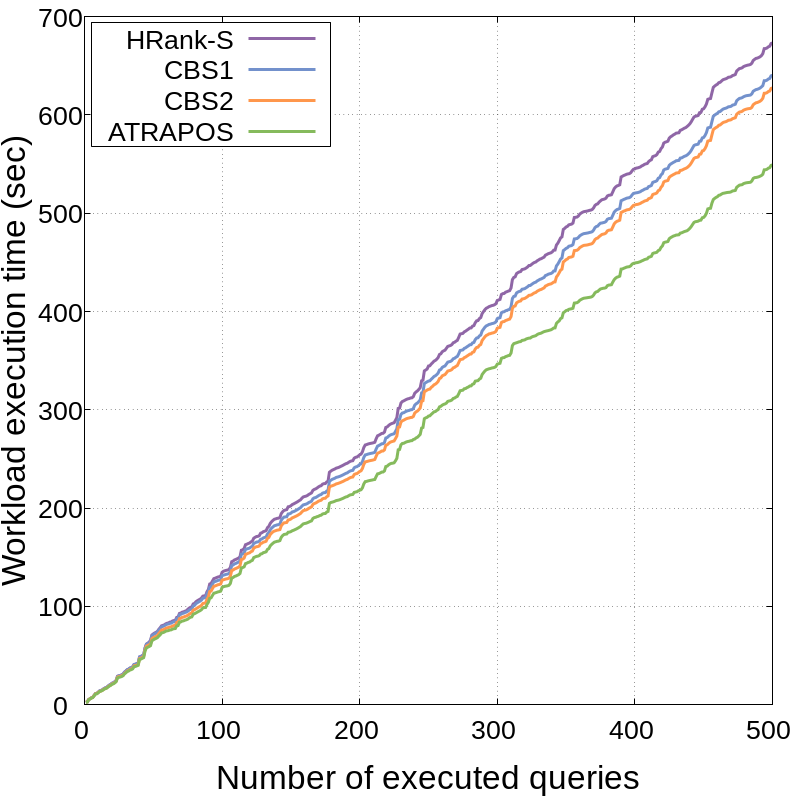}%
\caption{\gdelt}%
\label{subfigb}%
\end{subfigure}\hfill%
\caption{Evaluation against baseline caching approaches considering cumulative time per query.}
\label{fig:rivals-per-query-cumulative}
\end{figure}

\begin{figure}[t]
\centering
\begin{subfigure}{.5\columnwidth}
\centering
\includegraphics[width=.95\columnwidth]{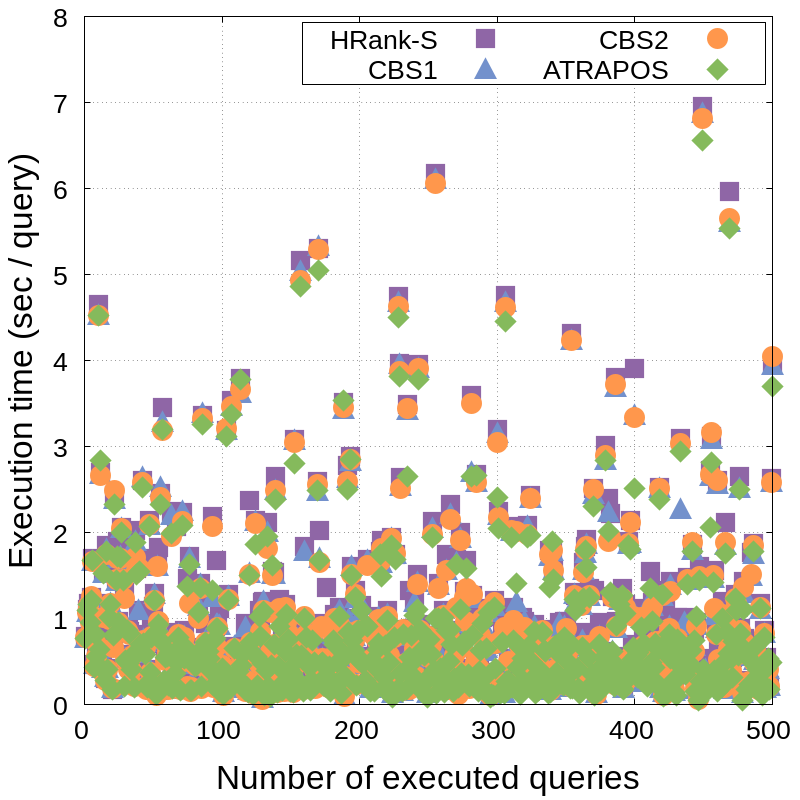}%
\caption{\dblp}%
\label{subfiga}%
\end{subfigure}\hfill%
\begin{subfigure}{.5\columnwidth}
\centering
\includegraphics[width=.95\columnwidth]{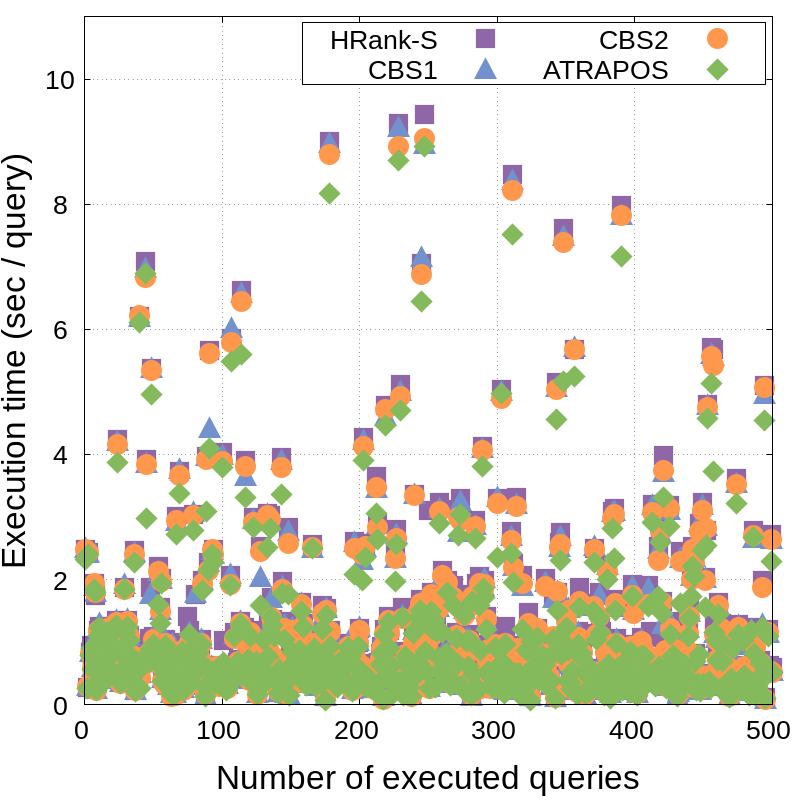}%
\caption{\gdelt}%
\label{subfigb}%
\end{subfigure}\hfill%
\centering
\begin{subfigure}{.5\columnwidth}
\centering
\includegraphics[width=.95\columnwidth]{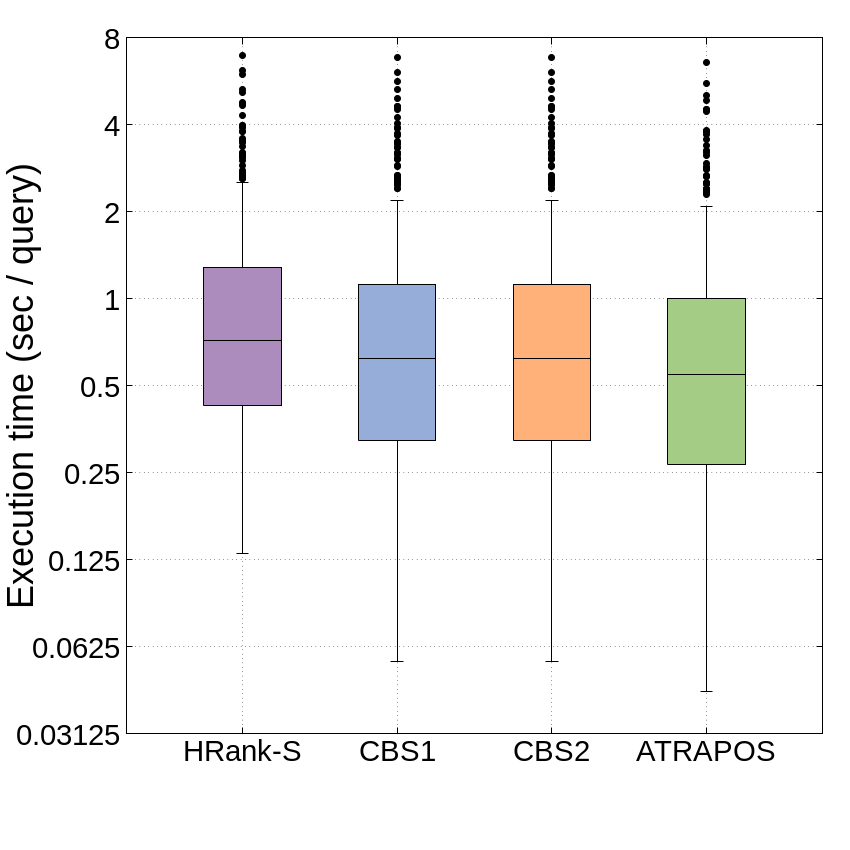}%
\caption{\dblp}%
\label{subfiga}%
\end{subfigure}\hfill%
\begin{subfigure}{.5\columnwidth}
\centering
\includegraphics[width=.95\columnwidth]{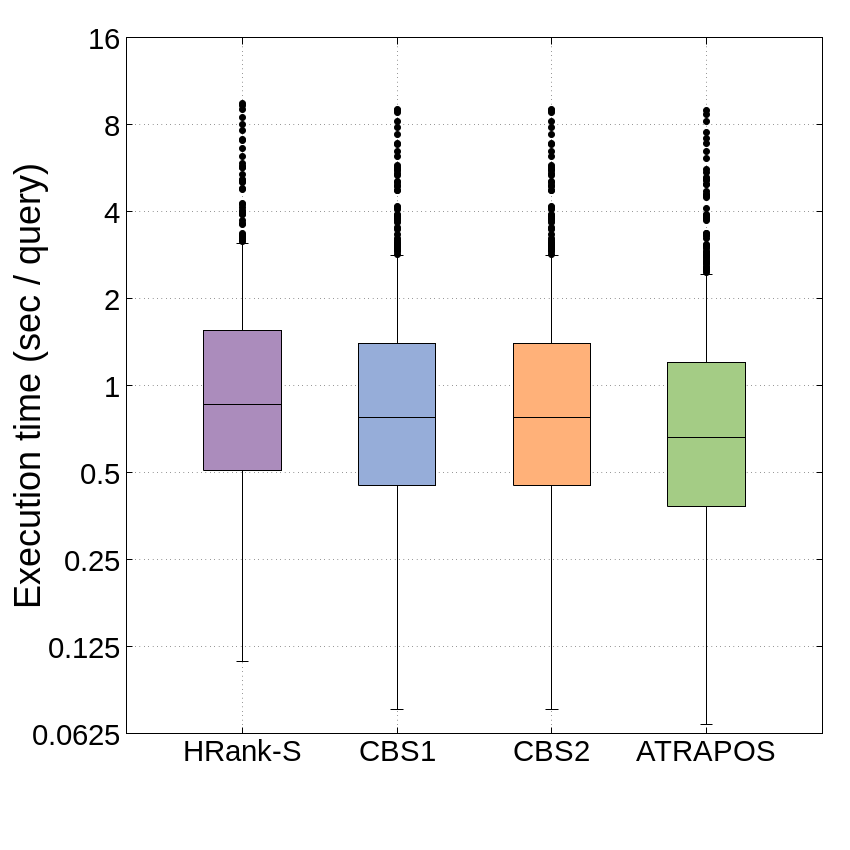}%
\caption{\gdelt}%
\label{subfigb}%
\end{subfigure}\hfill%
\caption{Evaluation against baseline caching approaches considering query execution time skewness.}
\label{fig:rivals-per-query}
\end{figure}

\subsubsection{Execution time per query}

In our previous experiments, we presented the average execution time per query across all query workloads. Here, we investigate the performance of individual queries, averaging the reported execution times by query position across~ $10$ workflows.
Figure~\ref{fig:rivals-per-query-cumulative}~shows the cumulative time during workload execution; 
cache-based approaches are noticeably faster than HRank-S, especially for the \dblp. 
\thiswork~is also considerably faster than CBS1 and CBS2 with the difference increasing with the number of queries.
Then, we investigate the skewness of the query execution times for the same workloads.
Specifically, Figures~\ref{fig:rivals-per-query}a-b illustrate the execution time per query; 
it is highlighted that \thiswork~achieves considerable gains in some of the slower queries.
Figures~\ref{fig:rivals-per-query}c-d reconfirm previous findings 
as each quartile of \thiswork~starts lower than the respective ones of the competitor approaches in both box plots.

\subsection{Effect of cache replacement policies}\label{sec:against_policies}

We now benchmark our cache replacement policy against the Least Recently Used (LRU)~\cite{aho1971} and Popularity-aware Greedy Dual-Size (PGDS)~\cite{Jin2000} policies (Section~\ref{sec:cache-manager}). 
We examine performance under varying cache size (Section~\ref{sec:against_policies-cache_size}), dataset size (Section~\ref{sec:against_policies-dataset_size}), session restart probability~$p$ (Section~\ref{sec:against_policies-sessions}) and distributions for selecting metapath queries (Section~\ref{sec:against_policies-zipf}). All methods utilise the Overlap Tree to identify query overlaps; their difference lies solely in the underlying cache replacement policy.

\begin{figure}[t]
\centering
\begin{subfigure}{.5\columnwidth}
\centering
\includegraphics[width=.95\columnwidth]{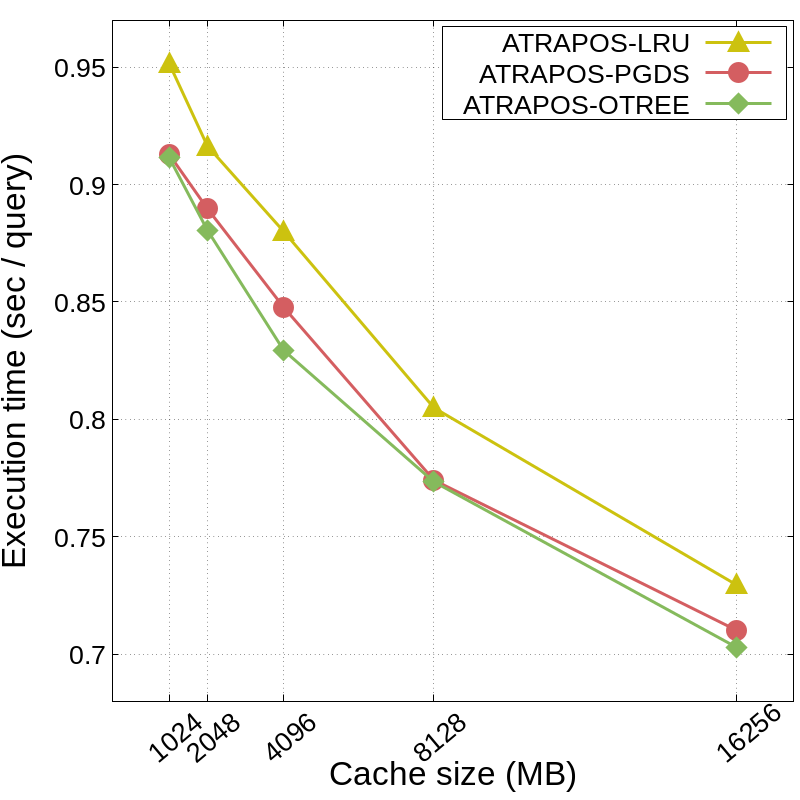}%
\caption{\dblp}%
\label{subfiga}%
\end{subfigure}\hfill%
\begin{subfigure}{.5\columnwidth}
\centering
\includegraphics[width=.95\columnwidth]{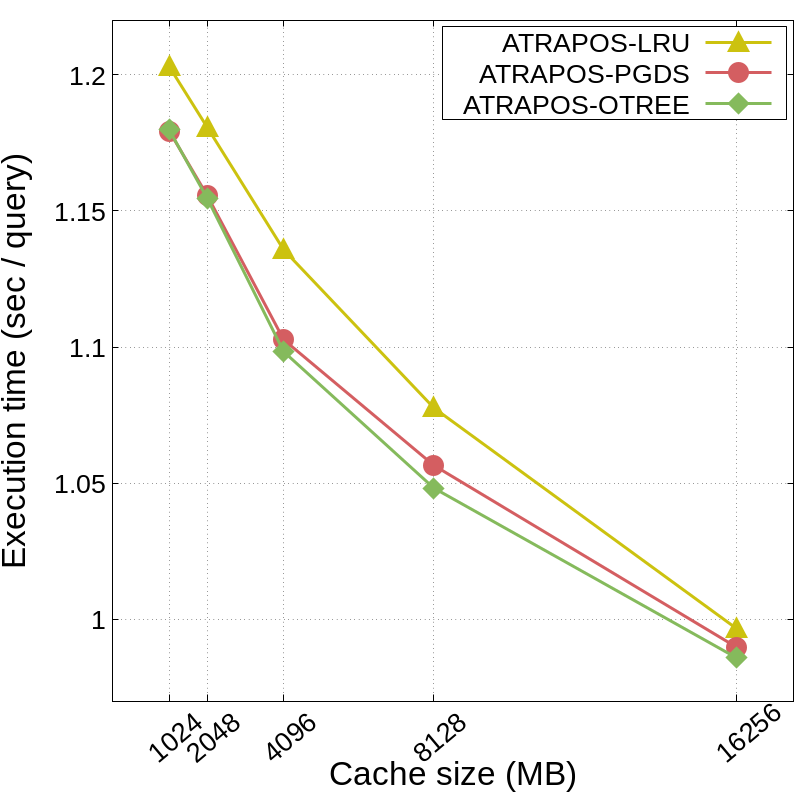}%
\caption{\gdelt}%
\label{subfigb}%
\end{subfigure}\hfill%
\caption{Evaluation against cache replacement policies with varying cache size.}
\label{fig:policies-cache}
\end{figure}

\subsubsection{Varying cache size.}\label{sec:against_policies-cache_size}

Figure~\ref{fig:policies-cache} shows our findings with varying cache size.
Standard \thiswork outperforms PGDS, the best performer among the rest on most scenarios; its advantage is more prominent on \dblp.
Yet, \thiswork is marginally faster than the PGDS policy on \gdelt; the most notable difference appears with cache size equal to $8$GB; at the same time, both cache policies incorporating frequency and item size outperform LRU.

\begin{figure}[t]
\begin{subfigure}{.5\columnwidth}
\centering
\includegraphics[width=.95\columnwidth]{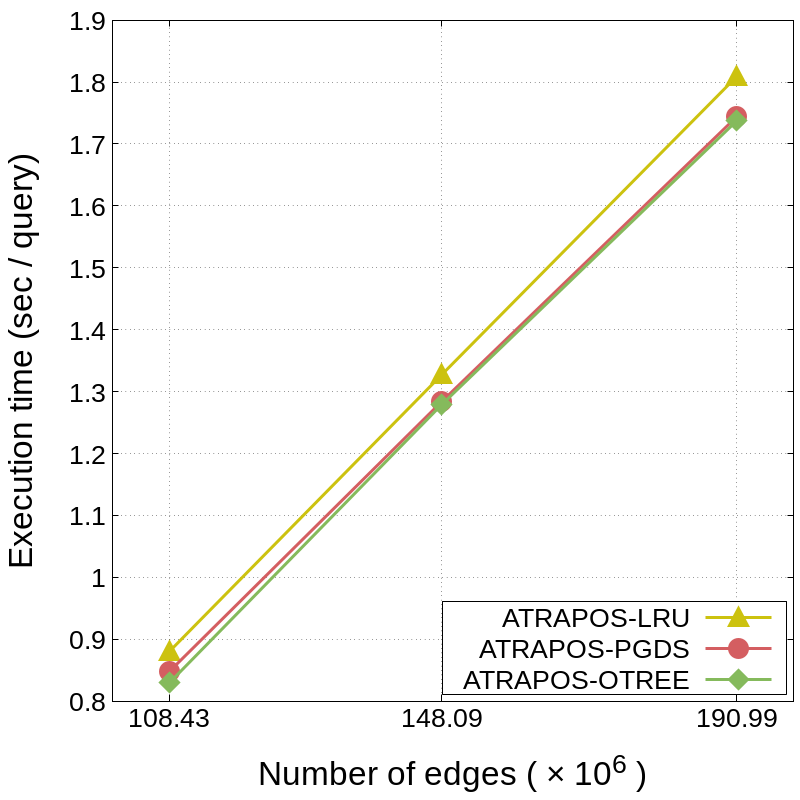}%
\caption{\dblp}
\label{subfiga}%
\end{subfigure}\hfill%
\begin{subfigure}{.5\columnwidth}
\centering
\includegraphics[width=.95\columnwidth]{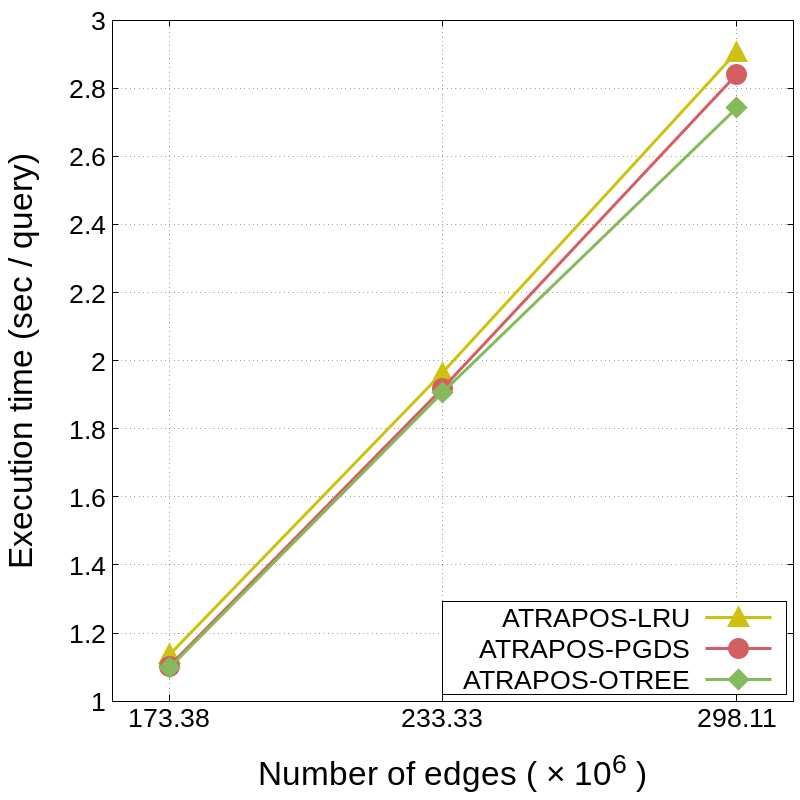}%
\caption{\gdelt}%
\label{subfigb}%
\end{subfigure}\hfill%
\caption{Evaluation against cache replacement policies with varying dataset size.}
\label{fig:policies-data-size}
\end{figure}

\subsubsection{Varying dataset size.}\label{sec:against_policies-dataset_size}

Figure~\ref{fig:policies-data-size} presents our results for all policies as we vary the dataset size. We observe a linear increase in execution time per query for all approaches as dataset size grows. LRU performs poorly compared to the other approaches in both datasets. \thiswork is faster than PGDS in most examined settings; it is marginally faster in the \dblp, with performance gains being  more apparent when using the full~\gdelt dataset that comprises~$298.11$ million edges.

\begin{figure}[t]
\centering
\begin{subfigure}{.5\columnwidth}
\centering
\includegraphics[width=.95\columnwidth]{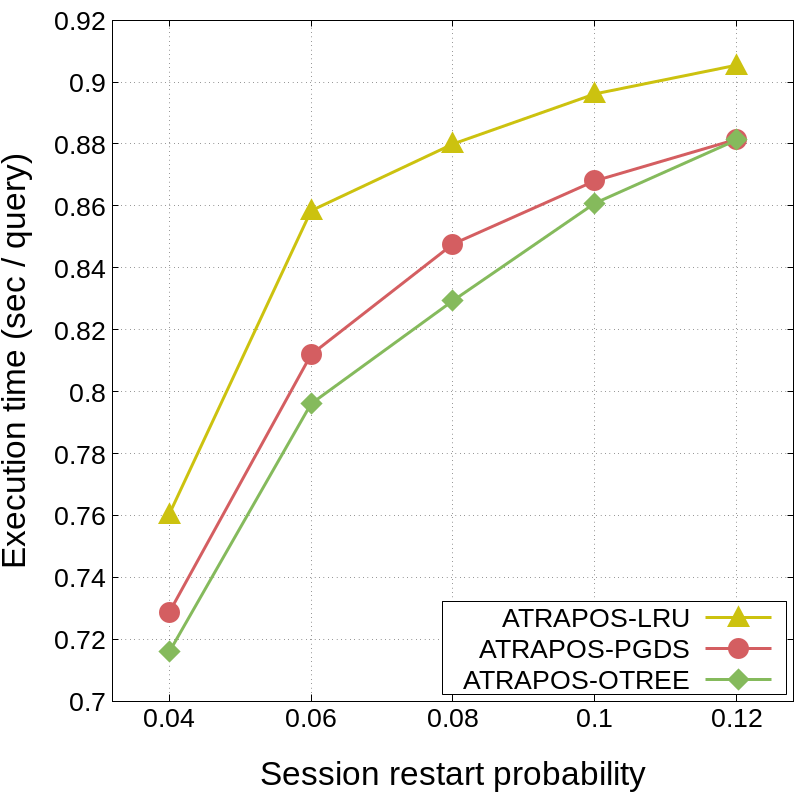}%
\caption{\dblp}%
\label{subfiga}%
\end{subfigure}\hfill%
\begin{subfigure}{.5\columnwidth}
\centering
\includegraphics[width=.95\columnwidth]{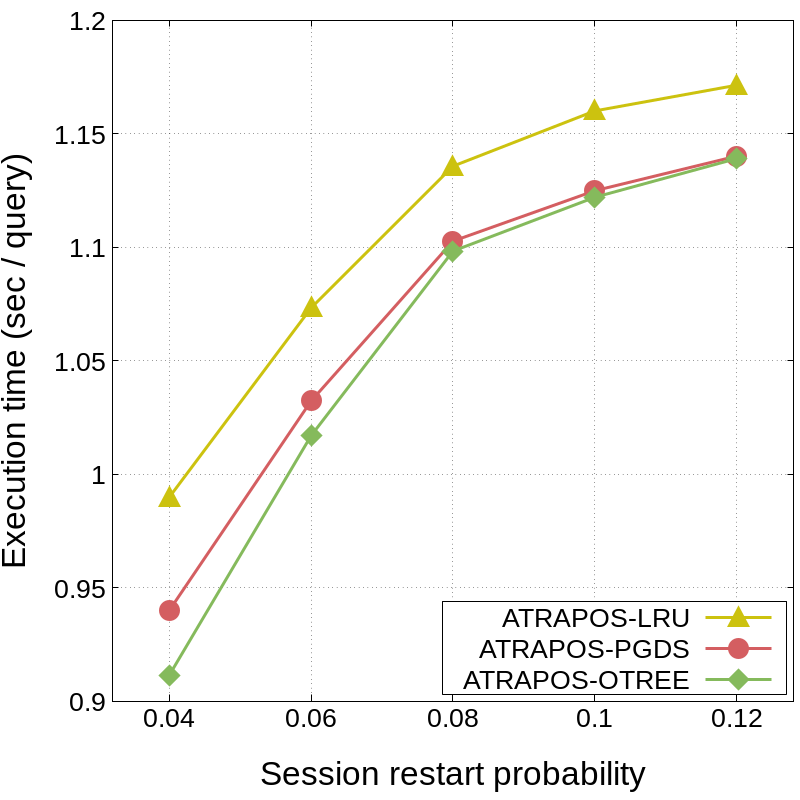}%
\caption{\gdelt}%
\label{subfigb}%
\end{subfigure}\hfill%
\caption{Evaluation against cache replacement policies varying session restart probability.}
\label{fig:policies-sessions}
\end{figure}

\subsubsection{Varying session restart probability $p$}\label{sec:against_policies-sessions}

Figure~\ref{fig:policies-sessions} presents performance when varying the session restart probability~$p$. The speedup falls for all methods as $p$ grows, which is reasonable, as a larger $p$ results in sessions containing fewer queries and thus fewer overlaps to exploit. \thiswork attains larger gains than other contestants in most examined configurations. More noteworthy differences appear on the \dblp, especially for~$p$ equal to $0.04$, $0.06$ and $0.08$. In the \gdelt, \thiswork~is faster for small values of $p$ i.e. $0.04$ and $0.06$, while PGDS achieves comparable results with \thiswork for larger~$p$. Both of them outperform LRU, with significant differences in both datasets.

\begin{figure}[t]
\begin{subfigure}{.5\columnwidth}
\centering
\includegraphics[width=.95\columnwidth]{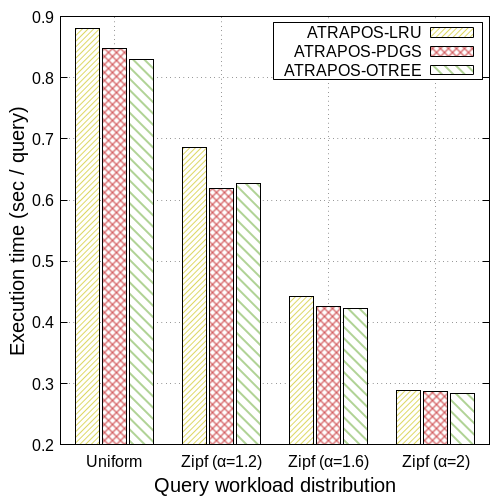}%
\caption{\dblp}%
\label{subfiga}%
\end{subfigure}\hfill%
\begin{subfigure}{.5\columnwidth}
\centering
\includegraphics[width=.95\columnwidth]{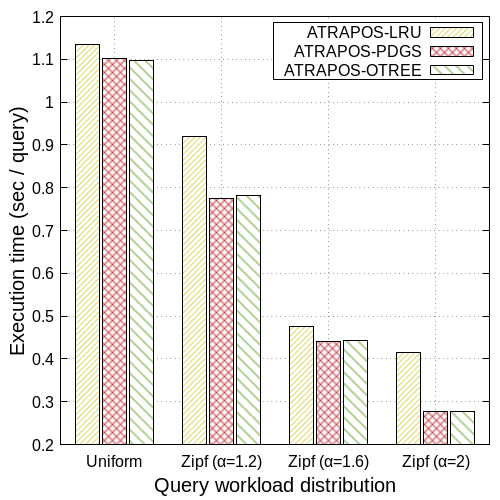}%
\caption{\gdelt}%
\label{subfigb}%
\end{subfigure}\hfill%
\caption{Evaluation against cache replacement policies with the queries following the Zipf distribution.}
\label{fig:policies-zipf}
\end{figure}

\subsubsection{Zipfian query workloads.}\label{sec:against_policies-zipf}

Figure~\ref{fig:policies-zipf} illustrates the performance of the examined cache replacement policies while using a Zipfian distribution for query workload generation. We observe a notable performance improvement with all approaches when generating the query workload by a Zipfian distribution compared to that when using a uniform distribution. Moreover, as parameter~$\alpha$ of the Zipfian distribution grows, the gains for all methods become more prominent. A larger~$\alpha$ indicates a higher probability for query repetitions in the workload. PGDS and \thiswork achieve comparable results on both datasets, with \thiswork being marginally faster with the uniform distribution and the Zipfian with~$\alpha$ equal to~$1.6$ and~$2$ on the \dblp. Still, both of these approaches outperform the LRU on both datasets.

\begin{figure}[t]
\begin{subfigure}{.5\columnwidth}
\centering
\includegraphics[width=.95\columnwidth]{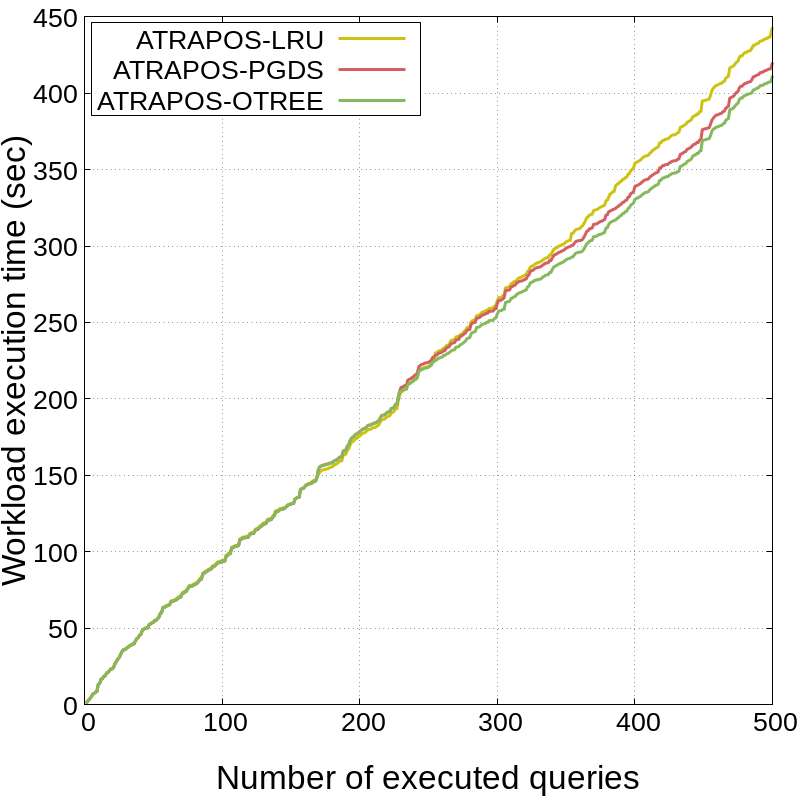}%
\caption{\dblp}%
\label{subfiga}%
\end{subfigure}\hfill%
\begin{subfigure}{.5\columnwidth}
\centering
\includegraphics[width=.95\columnwidth]{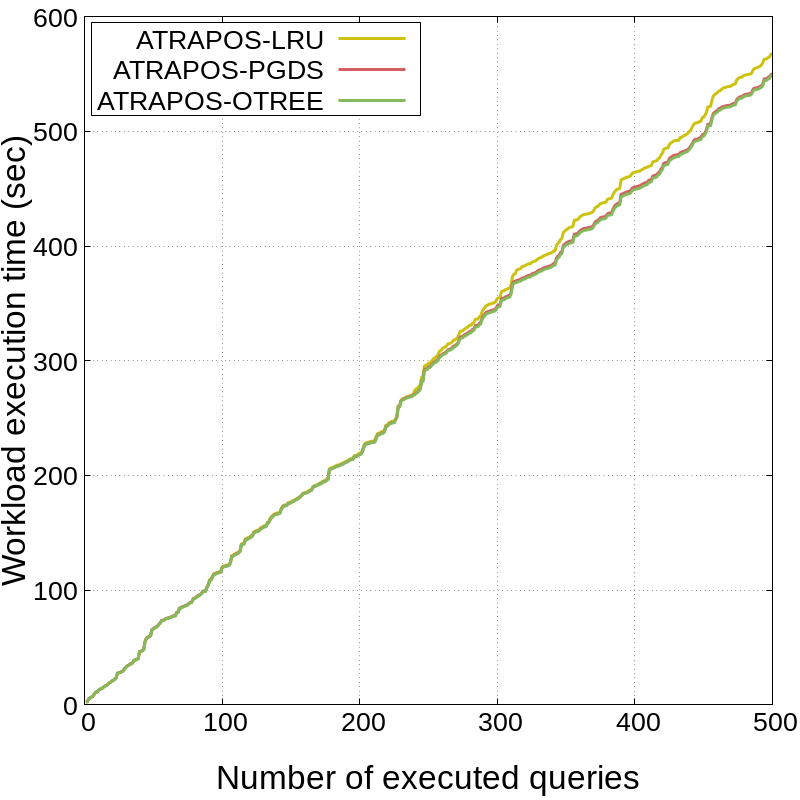}%
\caption{\gdelt}%
\label{subfigb}%
\end{subfigure}\hfill%
\caption{Evaluation against cache replacement policies considering cumulative time per query.}
\label{fig:policies-per-query-cumulative}
\end{figure}

\begin{figure}[t]
\centering
\begin{subfigure}{.5\columnwidth}
\centering
\includegraphics[width=.95\columnwidth]{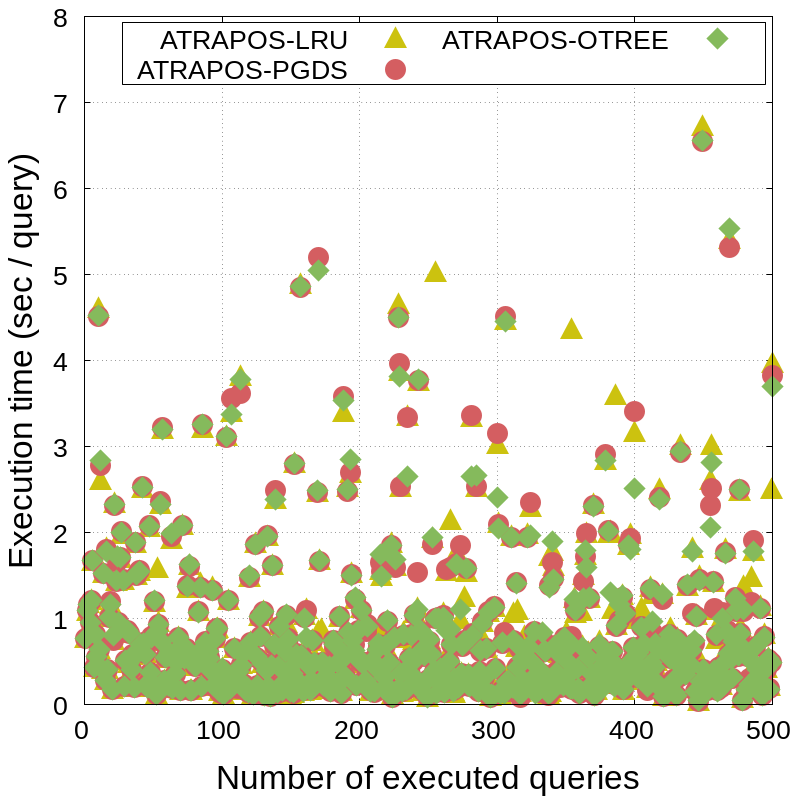}%
\caption{\dblp}%
\label{subfiga}%
\end{subfigure}\hfill%
\begin{subfigure}{.5\columnwidth}
\centering
\includegraphics[width=.95\columnwidth]{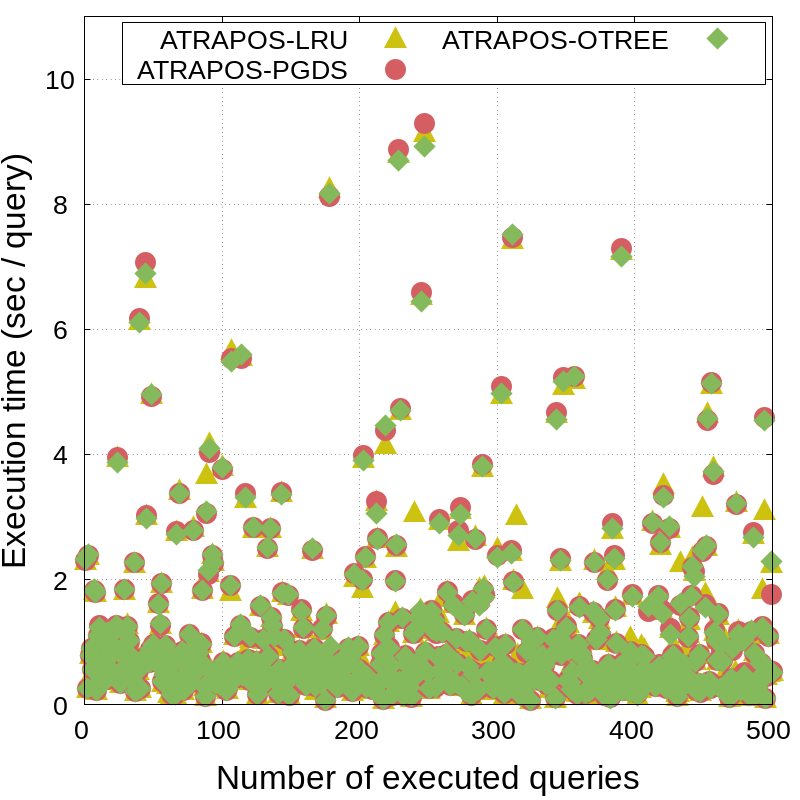}%
\caption{\gdelt}%
\label{subfigb}%
\end{subfigure}\hfill%
\centering
\begin{subfigure}{.5\columnwidth}
\centering
\includegraphics[width=.95\columnwidth]{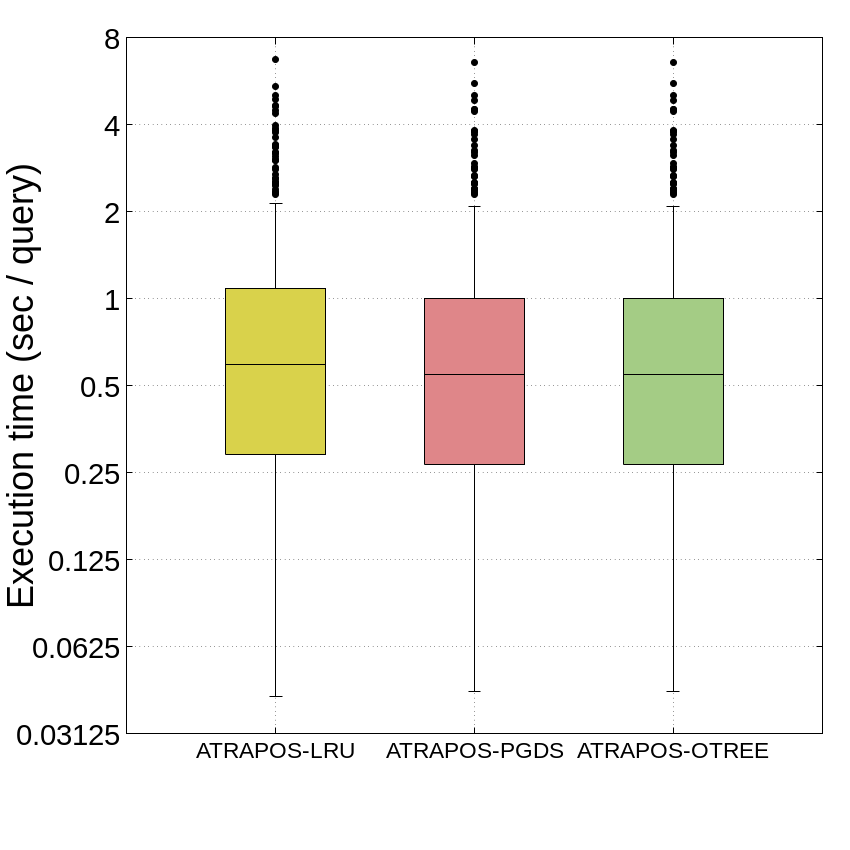}%
\caption{\dblp}%
\label{subfiga}%
\end{subfigure}\hfill%
\begin{subfigure}{.5\columnwidth}
\centering
\includegraphics[width=.95\columnwidth]{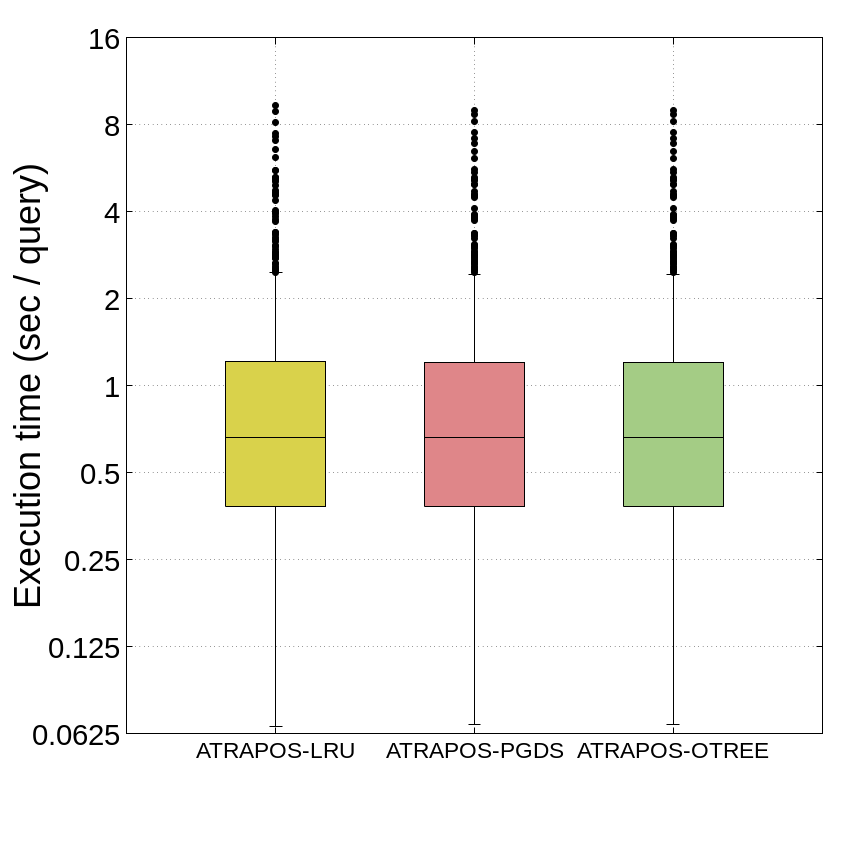}%
\caption{\gdelt}%
\label{subfigb}%
\end{subfigure}\hfill%
\caption{Evaluation against cache replacement policies considering query execution time skewness.}
\label{fig:policies-per-query}
\end{figure}

\subsubsection{Execution time per query.}\label{sec:against_policies-per-query}

Figure~\ref{fig:policies-per-query-cumulative} shows the cumulative times for the different cache replacement policies during workload execution.
\thiswork and PGDS achieve comparable times, yet \thiswork~is slightly faster on the \dblp, as evident in Figure~\ref{fig:policies-per-query-cumulative}a. 
LRU is slower than its competitors in both datasets; the difference is visible in the second half of both query workload executions.
Figures~\ref{fig:policies-per-query}a-b~plot the execution time of each metapath query for the same workloads. 
We discern some slower queries in the \dblp (Figure~\ref{fig:policies-per-query}a) in which LRU achieves considerably higher execution times that the other two approaches.
Last but not least, Figures~\ref{fig:policies-per-query}c reconfirms that PGDS and \thiswork~are faster than the LRU as each quartile in their box plots starts lower than the respective quartile for LRU for the \dblp. 
Of course, all considered approaches in this section utilise the Overlap Tree, differing only in terms of the cache replacement policy. Therefore, the difference in performance is small, confirming what someone would expect.

\section{Related Work}\label{sec:related}

Here, we review related work in HIN analysis, metapath query evaluation, multi-query optimisation, and cache replacement.

\subsection{HIN analysis}
Heterogeneous information networks (HINs) are subject to growing attention, as they afford a convenient model to represent information in several domains. Metapaths are instrumental in many data mining tasks in HINs and are widely used to capture complex semantics among HIN nodes. Several works rank HIN nodes based on their centrality in a metapath-defined network~\cite{multirank, li2014, pathrank}. HRank~\cite{hrank} co-ranks HIN nodes and relations based on the intuition that important nodes are connected to other nodes through important relations.
Other works propose metapath-based similarity measures among HIN nodes that encode different similarity semantics and are used in a variety of similarity search and join scenarios~\cite{sun11, Hetesim, shi2012, joinsim}. 
As different metapaths reveal different associations among HIN nodes, recent works incorporate metapath-based connectivity in link prediction and recommendation HIN analysis tasks~\cite{DBLP:conf/icdm/CaoKY14, DBLP:conf/kdd/ChenYWWNL18, DBLP:journals/tkde/ShiHZY19}.  
Furthermore, several community detection algorithms have been proposed for HINs; some approaches detect clusters that contain nodes of different types~\cite{shi2014ranking, sun2012relation, sun2009ranking}, while others generate homogeneous clusters~\cite{sun2009rankclus, sun2013pathselclus, zhou2013social}. In particular, Pathselclus~\cite{sun2009rankclus} uses user-defined metapaths as seeds and automatically adjusts their weights during the clustering process. 
As community detection algorithms take long time to form all communities for large networks,~\cite{fang2020effective} detects the community of a given node in a HIN extending traditional metrics for community cohesiveness to take metapaths into acccount.

\subsection{Metapath query evaluation}
A recent line of work focuses on metapath query evaluation. Shi et al.~\cite{hrank} apply dynamic programming~\cite{DynP} to reorganize the order of matrix multiplications required for metapath query evaluation, thereby reducing computational cost. When small errors can be tolerated, truncation strategies~\cite{Chakrabarti2007, Lao2010} can be applied in each step of matrix multiplication, zeroing values below a certain threshold. Another approach follows a Monte Carlo strategy~\cite{MC}~to approximate the query result via random walkers; the number of which strikes a balance in the tradeoff between accuracy and computation cost. Unlike \thiswork, these techniques focus on the performance of single-query evaluation, incognizant of data exploration workloads with naturally emerging overlapping queries. 
Finally, Sun et al.~\cite{sun11} 
propose the offline materialisation of short metapaths in 
a preprocessing stage and concatenate them at query time to make the online evaluation of a single query more efficient.

\subsection{Multi-query optimisation}
Multi-query optimisation (MQO) has been heavily studied, in particular for relational databases~\cite{finkelstein1982common, sellis1988multiple, sellis1990multiple, roy2000efficient, kathuria2017efficient}.
These works generate a globally optimal plan by computing common sub-expressions within query batches once and reuse them. In addition, 
MQO techniques have also been introduced in structural graph pattern search using graph isomorphism~\cite{ren2016multi}.
Although these works share the same principles as \thiswork, they do not focus on HINs while they greatly differ assuming queries are evaluated in batches;
this requires that query workloads are known a priori to determine common computations across queries, which is not the case in real-time applications. \thiswork, on the other hand, identifies query overlaps in real-time and automatically caches the respective results if it is beneficial (according to their occurrence frequency and their utility for the evaluation of other queries). Lastly, MQO techniques for RDF-triple stores have been proposed~\cite{papailiou2015graph, madkour2018worq}. However, apart from the fact that RDF graphs are less compact than HINs and hence likely to result in increased memory footprint and execution times, such works consider \emph{graph pattern queries} rather than metapaths; thus, \thiswork is customized for metapath-based HIN analysis.

\subsection{Cache replacement policies}
There is a vast literature on cache replacement, especially in web server caching~\cite{cache-survey-1, cache-survey-2}. Much of that work follows the Least Recently Used (LRU) policy~\cite{aho1971}; LRU-MIN~\cite{LRU_MIN} favors retaining smaller items in the cache, while LRU-S~\cite{LRU_S} incorporates the size of cached items into a randomized caching mechanism. Size-adjusted LRU (SLRU)\cite{SLRU}, considers both the size and cost of cached items. Greedy Dual-based approaches~\cite{GD, GDS} form another class of cache replacement policies. Notably, Popularity-aware GreedyDual-Size (PGDS)~\cite{Jin2000} utilizes the frequency, size, and cost of cached items. GreedyDual* (GD*)~\cite{GDStar} adaptively adjusts the significance of \emph{recency} and \emph{frequency} information over time. Other policies take into consideration the size and frequency of cached items~\cite{LRV, LV, LNC-R-W3-U}. An interesting approach~\cite{LUV} balances the recency and frequency of cached items with different costs with the help of a user-defined parameter. Nevertheless, none of the aforementioned policies considers the \emph{interdependence} among cached items that inevitably appears in a metapath query workload, as \thiswork does.

\section{Conclusions}\label{sec:conclusions}

We introduced \thiswork{}, a method to evaluate metapath query workloads in semantic knowledge graph and heterogeneous information network analysis tasks that exploits overlaps among query patterns. \thiswork{} accelerates metapath computations using matrix multiplication algorithms tailored for sparse matrices along with an appropriate cost model. Furthermore, it introduces an overlap-aware cache replacement policy leveraging a tailor-made data structure, the \emph{Overlap Tree}, that identifies overlapping metapaths among queries. Our thorough experimental evaluation on two large real-world HINs shows that \thiswork{} outperforms existing approaches in efficiency across all examined scenarios.

\begin{acks}
This work was partially funded by the EU H2020 project Smart-DataLake (GA: 825041)
and the EU Horizon Europe projects SciLake (GA: 101058573) and STELAR (GA: 101070122).
\end{acks}

\bibliographystyle{ACM-Reference-Format}
\bibliography{main}


\begin{thebibliography}{72}

%%% the Web address.

\ifx \showCODEN    \undefined \def \showCODEN     #1{\unskip}     \fi
\ifx \showDOI      \undefined \def \showDOI       #1{#1}\fi
\ifx \showISBNx    \undefined \def \showISBNx     #1{\unskip}     \fi
\ifx \showISBNxiii \undefined \def \showISBNxiii  #1{\unskip}     \fi
\ifx \showISSN     \undefined \def \showISSN      #1{\unskip}     \fi
\ifx \showLCCN     \undefined \def \showLCCN      #1{\unskip}     \fi
\ifx \shownote     \undefined \def \shownote      #1{#1}          \fi
\ifx \showarticletitle \undefined \def \showarticletitle #1{#1}   \fi
\ifx \showURL      \undefined \def \showURL       {\relax}        \fi
% invisible to TeX
\providecommand\bibfield[2]{#2}
\providecommand\bibinfo[2]{#2}
\providecommand\natexlab[1]{#1}
\providecommand\showeprint[2][]{arXiv:#2}

\bibitem[\protect\citeauthoryear{Abrams, Standridge, Abdulla, Williams, and
  Fox}{Abrams et~al\mbox{.}}{1995}]%
        {LRU_MIN}
\bibfield{author}{\bibinfo{person}{Marc Abrams}, \bibinfo{person}{Charles~R
  Standridge}, \bibinfo{person}{Ghaleb Abdulla}, \bibinfo{person}{Stephen
  Williams}, {and} \bibinfo{person}{Edward~A Fox}.}
  \bibinfo{year}{1995}\natexlab{}.
\newblock \bibinfo{booktitle}{\emph{Caching proxies: Limitations and
  potentials}}.
\newblock \bibinfo{type}{{T}echnical {R}eport}.
  \bibinfo{institution}{Department of Computer Science, Virginia Polytechnic
  Institute \& State}.
\newblock


\bibitem[\protect\citeauthoryear{Aggarwal, Wolf, and Yu}{Aggarwal
  et~al\mbox{.}}{1999}]%
        {SLRU}
\bibfield{author}{\bibinfo{person}{Charu Aggarwal}, \bibinfo{person}{Joel~L
  Wolf}, {and} \bibinfo{person}{Philip~S. Yu}.}
  \bibinfo{year}{1999}\natexlab{}.
\newblock \showarticletitle{Caching on the world wide web}.
\newblock \bibinfo{journal}{\emph{TKDE}} \bibinfo{volume}{11},
  \bibinfo{number}{1} (\bibinfo{year}{1999}), \bibinfo{pages}{94--107}.
\newblock


\bibitem[\protect\citeauthoryear{Aho, Denning, and Ullman}{Aho
  et~al\mbox{.}}{1971}]%
        {aho1971}
\bibfield{author}{\bibinfo{person}{Alfred~V Aho}, \bibinfo{person}{Peter~J
  Denning}, {and} \bibinfo{person}{Jeffrey~D Ullman}.}
  \bibinfo{year}{1971}\natexlab{}.
\newblock \showarticletitle{Principles of optimal page replacement}.
\newblock \bibinfo{journal}{\emph{JACM}} \bibinfo{volume}{18},
  \bibinfo{number}{1} (\bibinfo{year}{1971}), \bibinfo{pages}{80--93}.
\newblock


\bibitem[\protect\citeauthoryear{Asay}{Asay}{[n.d.]}]%
        {neo4jfund}
\bibfield{author}{\bibinfo{person}{Matt Asay}.}
  \bibinfo{year}{[n.d.]}\natexlab{}.
\newblock \bibinfo{booktitle}{\emph{Neo4j's \$325 million venture suggests
  databases are cool again}}.
\newblock
\urldef\tempurl%
\url{https://www.techrepublic.com/article/neo4js-325-million-venture-suggests-databases-are-cool-again/}
\showURL{%
\tempurl}


\bibitem[\protect\citeauthoryear{Bahn, Koh, Noh, and Lyul}{Bahn
  et~al\mbox{.}}{2002}]%
        {LUV}
\bibfield{author}{\bibinfo{person}{Hyokyung Bahn}, \bibinfo{person}{Kern Koh},
  \bibinfo{person}{Sam~H Noh}, {and} \bibinfo{person}{SM Lyul}.}
  \bibinfo{year}{2002}\natexlab{}.
\newblock \showarticletitle{Efficient replacement of nonuniform objects in web
  caches}.
\newblock \bibinfo{journal}{\emph{Computer}} \bibinfo{volume}{35},
  \bibinfo{number}{6} (\bibinfo{year}{2002}), \bibinfo{pages}{65--73}.
\newblock


\bibitem[\protect\citeauthoryear{Balamash and Krunz}{Balamash and
  Krunz}{2004}]%
        {cache-survey-2}
\bibfield{author}{\bibinfo{person}{Abdullah Balamash} {and}
  \bibinfo{person}{Marwan Krunz}.} \bibinfo{year}{2004}\natexlab{}.
\newblock \showarticletitle{An overview of web caching replacement algorithms}.
\newblock \bibinfo{journal}{\emph{IEEE Communications Surveys \& Tutorials}}
  \bibinfo{volume}{6}, \bibinfo{number}{2} (\bibinfo{year}{2004}),
  \bibinfo{pages}{44--56}.
\newblock


\bibitem[\protect\citeauthoryear{Bieganski, Riedl, Carlis, and
  Retzel}{Bieganski et~al\mbox{.}}{1994}]%
        {generalised_suffix_tree_2}
\bibfield{author}{\bibinfo{person}{Paul Bieganski}, \bibinfo{person}{John
  Riedl}, \bibinfo{person}{John~V Carlis}, {and} \bibinfo{person}{Ernest~F
  Retzel}.} \bibinfo{year}{1994}\natexlab{}.
\newblock \showarticletitle{Generalized suffix trees for biological sequence
  data: applications and implementation}. In \bibinfo{booktitle}{\emph{HICSS}}.
  \bibinfo{pages}{35--44}.
\newblock


\bibitem[\protect\citeauthoryear{Cao, Kong, and Yu}{Cao et~al\mbox{.}}{2014}]%
        {DBLP:conf/icdm/CaoKY14}
\bibfield{author}{\bibinfo{person}{Bokai Cao}, \bibinfo{person}{Xiangnan Kong},
  {and} \bibinfo{person}{Philip~S. Yu}.} \bibinfo{year}{2014}\natexlab{}.
\newblock \showarticletitle{Collective Prediction of Multiple Types of Links in
  Heterogeneous Information Networks}. In \bibinfo{booktitle}{\emph{{ICDM}}}.
  \bibinfo{pages}{50--59}.
\newblock


\bibitem[\protect\citeauthoryear{Cao and Irani}{Cao and Irani}{1997}]%
        {GDS}
\bibfield{author}{\bibinfo{person}{Pei Cao} {and} \bibinfo{person}{Sandy
  Irani}.} \bibinfo{year}{1997}\natexlab{}.
\newblock \showarticletitle{Cost-aware www proxy caching algorithms.}. In
  \bibinfo{booktitle}{\emph{Usenix}}, Vol.~\bibinfo{volume}{12}.
  \bibinfo{pages}{193--206}.
\newblock


\bibitem[\protect\citeauthoryear{Chakrabarti}{Chakrabarti}{2007}]%
        {Chakrabarti2007}
\bibfield{author}{\bibinfo{person}{Soumen Chakrabarti}.}
  \bibinfo{year}{2007}\natexlab{}.
\newblock \showarticletitle{Dynamic personalized pagerank in entity-relation
  graphs}. In \bibinfo{booktitle}{\emph{WWW}}. \bibinfo{pages}{571--580}.
\newblock


\bibitem[\protect\citeauthoryear{Chatzopoulos, Patroumpas, Zeakis, Vergoulis,
  and Skoutas}{Chatzopoulos et~al\mbox{.}}{2020}]%
        {chatzopoulos2020sphinx}
\bibfield{author}{\bibinfo{person}{Serafeim Chatzopoulos},
  \bibinfo{person}{Kostas Patroumpas}, \bibinfo{person}{Alexandros Zeakis},
  \bibinfo{person}{Thanasis Vergoulis}, {and} \bibinfo{person}{Dimitrios
  Skoutas}.} \bibinfo{year}{2020}\natexlab{}.
\newblock \showarticletitle{SPHINX: a system for metapath-based entity
  exploration in heterogeneous information networks}.
\newblock \bibinfo{journal}{\emph{VLDB}} \bibinfo{volume}{13},
  \bibinfo{number}{12} (\bibinfo{year}{2020}), \bibinfo{pages}{2913--2916}.
\newblock


\bibitem[\protect\citeauthoryear{Chatzopoulos, Vergoulis, Deligiannis, Skoutas,
  Dalamagas, and Tryfonopoulos}{Chatzopoulos et~al\mbox{.}}{2021}]%
        {chatzopoulos2021scinem}
\bibfield{author}{\bibinfo{person}{Serafeim Chatzopoulos},
  \bibinfo{person}{Thanasis Vergoulis}, \bibinfo{person}{Panagiotis
  Deligiannis}, \bibinfo{person}{Dimitrios Skoutas}, \bibinfo{person}{Theodore
  Dalamagas}, {and} \bibinfo{person}{Christos Tryfonopoulos}.}
  \bibinfo{year}{2021}\natexlab{}.
\newblock \showarticletitle{SciNeM: A Scalable Data Science Tool for
  Heterogeneous Network Mining.}. In \bibinfo{booktitle}{\emph{EDBT}}.
  \bibinfo{pages}{654--657}.
\newblock


\bibitem[\protect\citeauthoryear{Chen, Yin, Wang, Wang, Nguyen, and Li}{Chen
  et~al\mbox{.}}{2018}]%
        {DBLP:conf/kdd/ChenYWWNL18}
\bibfield{author}{\bibinfo{person}{Hongxu Chen}, \bibinfo{person}{Hongzhi Yin},
  \bibinfo{person}{Weiqing Wang}, \bibinfo{person}{Hao Wang},
  \bibinfo{person}{Quoc Viet~Hung Nguyen}, {and} \bibinfo{person}{Xue Li}.}
  \bibinfo{year}{2018}\natexlab{}.
\newblock \showarticletitle{{PME:} Projected Metric Embedding on Heterogeneous
  Networks for Link Prediction}. In \bibinfo{booktitle}{\emph{{SIGKDD}}}.
  \bibinfo{pages}{1177--1186}.
\newblock


\bibitem[\protect\citeauthoryear{Chen, Shuai, Yang, Lee, Shi, Philip, and
  Chen}{Chen et~al\mbox{.}}{2021}]%
        {chen2021structure}
\bibfield{author}{\bibinfo{person}{Hsi-Wen Chen}, \bibinfo{person}{Hong-Han
  Shuai}, \bibinfo{person}{De-Nian Yang}, \bibinfo{person}{Wang-Chien Lee},
  \bibinfo{person}{Chuan Shi}, \bibinfo{person}{S~Yu Philip}, {and}
  \bibinfo{person}{Ming-Syan Chen}.} \bibinfo{year}{2021}\natexlab{}.
\newblock \showarticletitle{Structure-Aware Parameter-Free Group Query via
  Heterogeneous Information Network Transformer}. In
  \bibinfo{booktitle}{\emph{ICDE}}. \bibinfo{pages}{2075--2080}.
\newblock


\bibitem[\protect\citeauthoryear{Cohen}{Cohen}{1998}]%
        {CohenEst}
\bibfield{author}{\bibinfo{person}{Edith Cohen}.}
  \bibinfo{year}{1998}\natexlab{}.
\newblock \showarticletitle{Structure Prediction and Computation of Sparse
  Matrix Products}.
\newblock \bibinfo{journal}{\emph{Journal of Combinatorial Optimization}}
  \bibinfo{volume}{2}, \bibinfo{number}{4} (\bibinfo{year}{1998}),
  \bibinfo{pages}{307--332}.
\newblock


\bibitem[\protect\citeauthoryear{Cormen, Leiserson, Rivest, and Stein}{Cormen
  et~al\mbox{.}}{2009}]%
        {DynP}
\bibfield{author}{\bibinfo{person}{Thomas~H. Cormen},
  \bibinfo{person}{Charles~E. Leiserson}, \bibinfo{person}{Ronald~L. Rivest},
  {and} \bibinfo{person}{Clifford Stein}.} \bibinfo{year}{2009}\natexlab{}.
\newblock \bibinfo{booktitle}{\emph{Introduction to Algorithms, 3rd Edition}}.
\newblock \bibinfo{publisher}{{MIT} Press}.
\newblock
\showISBNx{978-0-262-03384-8}
\urldef\tempurl%
\url{http://mitpress.mit.edu/books/introduction-algorithms}
\showURL{%
\tempurl}


\bibitem[\protect\citeauthoryear{Fang, Wang, Lin, and Zhang}{Fang
  et~al\mbox{.}}{2021}]%
        {fang2021cohesive}
\bibfield{author}{\bibinfo{person}{Yixiang Fang}, \bibinfo{person}{Kai Wang},
  \bibinfo{person}{Xuemin Lin}, {and} \bibinfo{person}{Wenjie Zhang}.}
  \bibinfo{year}{2021}\natexlab{}.
\newblock \showarticletitle{Cohesive Subgraph Search over Big Heterogeneous
  Information Networks: Applications, Challenges, and Solutions}. In
  \bibinfo{booktitle}{\emph{SIGMOD}}. \bibinfo{pages}{2829--2838}.
\newblock


\bibitem[\protect\citeauthoryear{Fang, Yang, Zhang, Lin, and Cao}{Fang
  et~al\mbox{.}}{2020}]%
        {fang2020effective}
\bibfield{author}{\bibinfo{person}{Yixiang Fang}, \bibinfo{person}{Yixing
  Yang}, \bibinfo{person}{Wenjie Zhang}, \bibinfo{person}{Xuemin Lin}, {and}
  \bibinfo{person}{Xin Cao}.} \bibinfo{year}{2020}\natexlab{}.
\newblock \showarticletitle{Effective and efficient community search over large
  heterogeneous information networks}.
\newblock \bibinfo{journal}{\emph{VLDB}} \bibinfo{volume}{13},
  \bibinfo{number}{6} (\bibinfo{year}{2020}), \bibinfo{pages}{854--867}.
\newblock


\bibitem[\protect\citeauthoryear{Finkelstein}{Finkelstein}{1982}]%
        {finkelstein1982common}
\bibfield{author}{\bibinfo{person}{Sheldon Finkelstein}.}
  \bibinfo{year}{1982}\natexlab{}.
\newblock \showarticletitle{Common expression analysis in database
  applications}. In \bibinfo{booktitle}{\emph{SIGMOD}}.
  \bibinfo{pages}{235--245}.
\newblock


\bibitem[\protect\citeauthoryear{Foong, Hu, and Heisey}{Foong
  et~al\mbox{.}}{1999}]%
        {LV}
\bibfield{author}{\bibinfo{person}{Annie~P Foong}, \bibinfo{person}{Yu-Hen Hu},
  {and} \bibinfo{person}{Dennis~M Heisey}.} \bibinfo{year}{1999}\natexlab{}.
\newblock \showarticletitle{Adaptive web caching using logistic regression}. In
  \bibinfo{booktitle}{\emph{SSP}}. \bibinfo{pages}{515--524}.
\newblock


\bibitem[\protect\citeauthoryear{Guennebaud, Jacob, et~al\mbox{.}}{Guennebaud
  et~al\mbox{.}}{2010}]%
        {eigenweb}
\bibfield{author}{\bibinfo{person}{Ga\"{e}l Guennebaud},
  \bibinfo{person}{Beno\^{i}t Jacob}, {et~al\mbox{.}}}
  \bibinfo{year}{2010}\natexlab{}.
\newblock \bibinfo{title}{Eigen v3}.
\newblock \bibinfo{howpublished}{http://eigen.tuxfamily.org}.
\newblock


\bibitem[\protect\citeauthoryear{Jian, Wang, and Chen}{Jian
  et~al\mbox{.}}{2020}]%
        {jian2020effective}
\bibfield{author}{\bibinfo{person}{Xun Jian}, \bibinfo{person}{Yue Wang}, {and}
  \bibinfo{person}{Lei Chen}.} \bibinfo{year}{2020}\natexlab{}.
\newblock \showarticletitle{Effective and efficient relational community
  detection and search in large dynamic heterogeneous information networks}.
\newblock \bibinfo{journal}{\emph{VLDB}} \bibinfo{volume}{13},
  \bibinfo{number}{10} (\bibinfo{year}{2020}), \bibinfo{pages}{1723--1736}.
\newblock


\bibitem[\protect\citeauthoryear{Jin and Bestavros}{Jin and Bestavros}{2001}]%
        {GDStar}
\bibfield{author}{\bibinfo{person}{Shudong Jin} {and} \bibinfo{person}{Azer
  Bestavros}.} \bibinfo{year}{2001}\natexlab{}.
\newblock \showarticletitle{GreedyDual* Web caching algorithm: exploiting the
  two sources of temporal locality in Web request streams}.
\newblock \bibinfo{journal}{\emph{Computer Communications}}
  \bibinfo{volume}{24}, \bibinfo{number}{2} (\bibinfo{year}{2001}),
  \bibinfo{pages}{174--183}.
\newblock


\bibitem[\protect\citeauthoryear{Kathuria and Sudarshan}{Kathuria and
  Sudarshan}{2017}]%
        {kathuria2017efficient}
\bibfield{author}{\bibinfo{person}{Tarun Kathuria} {and} \bibinfo{person}{S
  Sudarshan}.} \bibinfo{year}{2017}\natexlab{}.
\newblock \showarticletitle{Efficient and provable multi-query optimization}.
  In \bibinfo{booktitle}{\emph{Symposium on Principles of Database Systems (ACM
  SIGMOD-SIGACT-SIGAI)}}. \bibinfo{pages}{53--67}.
\newblock


\bibitem[\protect\citeauthoryear{Kernert, K{\"o}hler, and Lehner}{Kernert
  et~al\mbox{.}}{2015}]%
        {SpMacho}
\bibfield{author}{\bibinfo{person}{David Kernert}, \bibinfo{person}{Frank
  K{\"o}hler}, {and} \bibinfo{person}{Wolfgang Lehner}.}
  \bibinfo{year}{2015}\natexlab{}.
\newblock \showarticletitle{SpMacho - Optimizing Sparse Linear Algebra
  Expressions with Probabilistic Density Estimation}. In
  \bibinfo{booktitle}{\emph{EDBT}}.
\newblock


\bibitem[\protect\citeauthoryear{Lao and Cohen}{Lao and Cohen}{2010}]%
        {Lao2010}
\bibfield{author}{\bibinfo{person}{Ni Lao} {and} \bibinfo{person}{William~W
  Cohen}.} \bibinfo{year}{2010}\natexlab{}.
\newblock \showarticletitle{Fast query execution for retrieval models based on
  path-constrained random walks}. In \bibinfo{booktitle}{\emph{SIGKDD}}.
  \bibinfo{pages}{881--888}.
\newblock


\bibitem[\protect\citeauthoryear{Lee, Park, Kahng, and Lee}{Lee
  et~al\mbox{.}}{2012}]%
        {pathrank}
\bibfield{author}{\bibinfo{person}{Sangkeun Lee}, \bibinfo{person}{Sungchan
  Park}, \bibinfo{person}{Minsuk Kahng}, {and} \bibinfo{person}{Sang~Goo Lee}.}
  \bibinfo{year}{2012}\natexlab{}.
\newblock \showarticletitle{PathRank: a novel node ranking measure on a
  heterogeneous graph for recommender systems}.
\newblock \bibinfo{journal}{\emph{CIKM}}.
\newblock


\bibitem[\protect\citeauthoryear{Leetaru and Schrodt}{Leetaru and
  Schrodt}{2013}]%
        {Leetaru2013}
\bibfield{author}{\bibinfo{person}{Kalev Leetaru} {and}
  \bibinfo{person}{Philip~A Schrodt}.} \bibinfo{year}{2013}\natexlab{}.
\newblock \showarticletitle{Gdelt: Global data on events, location, and tone,
  1979--2012}. In \bibinfo{booktitle}{\emph{ISA annual convention}},
  Vol.~\bibinfo{volume}{2}. Citeseer, \bibinfo{pages}{1--49}.
\newblock


\bibitem[\protect\citeauthoryear{Li, Ding, Kao, Sun, and Mamoulis}{Li
  et~al\mbox{.}}{2021}]%
        {li2021leveraging}
\bibfield{author}{\bibinfo{person}{Xiang Li}, \bibinfo{person}{Danhao Ding},
  \bibinfo{person}{Ben Kao}, \bibinfo{person}{Yizhou Sun}, {and}
  \bibinfo{person}{Nikos Mamoulis}.} \bibinfo{year}{2021}\natexlab{}.
\newblock \showarticletitle{Leveraging Meta-path Contexts for Classification in
  Heterogeneous Information Networks}. In \bibinfo{booktitle}{\emph{ICDE}}.
  \bibinfo{pages}{912--923}.
\newblock


\bibitem[\protect\citeauthoryear{Li, Shi, Philip, and Chen}{Li
  et~al\mbox{.}}{2014}]%
        {li2014}
\bibfield{author}{\bibinfo{person}{Yitong Li}, \bibinfo{person}{Chuan Shi},
  \bibinfo{person}{S~Yu Philip}, {and} \bibinfo{person}{Qing Chen}.}
  \bibinfo{year}{2014}\natexlab{}.
\newblock \showarticletitle{Hrank: a path based ranking method in heterogeneous
  information network}. In \bibinfo{booktitle}{\emph{WAIM}}.
  \bibinfo{pages}{553--565}.
\newblock


\bibitem[\protect\citeauthoryear{Liu}{Liu}{2008}]%
        {MC}
\bibfield{author}{\bibinfo{person}{Jun~S Liu}.}
  \bibinfo{year}{2008}\natexlab{}.
\newblock \bibinfo{booktitle}{\emph{Monte Carlo strategies in scientific
  computing}}.
\newblock \bibinfo{publisher}{Springer Science \& Business Media}.
\newblock


\bibitem[\protect\citeauthoryear{Madkour, Aly, and Aref}{Madkour
  et~al\mbox{.}}{2018}]%
        {madkour2018worq}
\bibfield{author}{\bibinfo{person}{Amgad Madkour}, \bibinfo{person}{Ahmed~M
  Aly}, {and} \bibinfo{person}{Walid~G Aref}.} \bibinfo{year}{2018}\natexlab{}.
\newblock \showarticletitle{Worq: Workload-driven rdf query processing}. In
  \bibinfo{booktitle}{\emph{International Semantic Web Conference (ISWC)}}.
  \bibinfo{pages}{583--599}.
\newblock


\bibitem[\protect\citeauthoryear{Meng, Cheng, Maniu, Senellart, and Zhang}{Meng
  et~al\mbox{.}}{2015}]%
        {DBLP:conf/www/MengCMSZ15}
\bibfield{author}{\bibinfo{person}{Changping Meng}, \bibinfo{person}{Reynold
  Cheng}, \bibinfo{person}{Silviu Maniu}, \bibinfo{person}{Pierre Senellart},
  {and} \bibinfo{person}{Wangda Zhang}.} \bibinfo{year}{2015}\natexlab{}.
\newblock \showarticletitle{Discovering Meta-Paths in Large Heterogeneous
  Information Networks}. In \bibinfo{booktitle}{\emph{{WWW}}}.
  \bibinfo{pages}{754--764}.
\newblock


\bibitem[\protect\citeauthoryear{Neumann and Weikum}{Neumann and
  Weikum}{2010}]%
        {rdf3x}
\bibfield{author}{\bibinfo{person}{Thomas Neumann} {and}
  \bibinfo{person}{Gerhard Weikum}.} \bibinfo{year}{2010}\natexlab{}.
\newblock \showarticletitle{The RDF-3X engine for scalable management of RDF
  data}.
\newblock \bibinfo{journal}{\emph{VLDB}} \bibinfo{volume}{19},
  \bibinfo{number}{1} (\bibinfo{year}{2010}), \bibinfo{pages}{91--113}.
\newblock


\bibitem[\protect\citeauthoryear{Ng, Li, and Ye}{Ng et~al\mbox{.}}{2011}]%
        {multirank}
\bibfield{author}{\bibinfo{person}{Michaek Kwok-Po Ng}, \bibinfo{person}{Xutao
  Li}, {and} \bibinfo{person}{Yunming Ye}.} \bibinfo{year}{2011}\natexlab{}.
\newblock \showarticletitle{Multirank: co-ranking for objects and relations in
  multi-relational data}. In \bibinfo{booktitle}{\emph{SIGKDD}}.
  \bibinfo{pages}{1217--1225}.
\newblock


\bibitem[\protect\citeauthoryear{Papailiou, Tsoumakos, Karras, and
  Koziris}{Papailiou et~al\mbox{.}}{2015}]%
        {papailiou2015graph}
\bibfield{author}{\bibinfo{person}{Nikolaos Papailiou},
  \bibinfo{person}{Dimitrios Tsoumakos}, \bibinfo{person}{Panagiotis Karras},
  {and} \bibinfo{person}{Nectarios Koziris}.} \bibinfo{year}{2015}\natexlab{}.
\newblock \showarticletitle{Graph-aware, workload-adaptive SPARQL query
  caching}. In \bibinfo{booktitle}{\emph{SIGMOD}}. \bibinfo{pages}{1777--1792}.
\newblock


\bibitem[\protect\citeauthoryear{Podlipnig and B{\"o}sz{\"o}rmenyi}{Podlipnig
  and B{\"o}sz{\"o}rmenyi}{2003}]%
        {cache-survey-1}
\bibfield{author}{\bibinfo{person}{Stefan Podlipnig} {and}
  \bibinfo{person}{Laszlo B{\"o}sz{\"o}rmenyi}.}
  \bibinfo{year}{2003}\natexlab{}.
\newblock \showarticletitle{A survey of web cache replacement strategies}.
\newblock \bibinfo{journal}{\emph{CSUR}} \bibinfo{volume}{35},
  \bibinfo{number}{4} (\bibinfo{year}{2003}), \bibinfo{pages}{374--398}.
\newblock


\bibitem[\protect\citeauthoryear{Ren and Wang}{Ren and Wang}{2016}]%
        {ren2016multi}
\bibfield{author}{\bibinfo{person}{Xuguang Ren} {and} \bibinfo{person}{Junhu
  Wang}.} \bibinfo{year}{2016}\natexlab{}.
\newblock \showarticletitle{Multi-query optimization for subgraph isomorphism
  search}.
\newblock \bibinfo{journal}{\emph{VLDB}} \bibinfo{volume}{10},
  \bibinfo{number}{3} (\bibinfo{year}{2016}), \bibinfo{pages}{121--132}.
\newblock


\bibitem[\protect\citeauthoryear{Rizzo and Vicisano}{Rizzo and
  Vicisano}{2000}]%
        {LRV}
\bibfield{author}{\bibinfo{person}{Luigi Rizzo} {and} \bibinfo{person}{Lorenzo
  Vicisano}.} \bibinfo{year}{2000}\natexlab{}.
\newblock \showarticletitle{Replacement policies for a proxy cache}.
\newblock \bibinfo{journal}{\emph{IEEE/ACM Transactions on Networking}}
  \bibinfo{volume}{8}, \bibinfo{number}{2} (\bibinfo{year}{2000}),
  \bibinfo{pages}{158--170}.
\newblock


\bibitem[\protect\citeauthoryear{Roy, Seshadri, Sudarshan, and Bhobe}{Roy
  et~al\mbox{.}}{2000}]%
        {roy2000efficient}
\bibfield{author}{\bibinfo{person}{Prasan Roy}, \bibinfo{person}{Srinivasan
  Seshadri}, \bibinfo{person}{S Sudarshan}, {and} \bibinfo{person}{Siddhesh
  Bhobe}.} \bibinfo{year}{2000}\natexlab{}.
\newblock \showarticletitle{Efficient and extensible algorithms for multi query
  optimization}. In \bibinfo{booktitle}{\emph{SIGMOD}}.
  \bibinfo{pages}{249--260}.
\newblock


\bibitem[\protect\citeauthoryear{Sellis and Ghosh}{Sellis and Ghosh}{1990}]%
        {sellis1990multiple}
\bibfield{author}{\bibinfo{person}{Timos Sellis} {and} \bibinfo{person}{Subrata
  Ghosh}.} \bibinfo{year}{1990}\natexlab{}.
\newblock \showarticletitle{On the multiple-query optimization problem}.
\newblock \bibinfo{journal}{\emph{Transactions on Knowledge and Data
  Engineering (TKDE)}} \bibinfo{volume}{2}, \bibinfo{number}{02}
  (\bibinfo{year}{1990}), \bibinfo{pages}{262--266}.
\newblock


\bibitem[\protect\citeauthoryear{Sellis}{Sellis}{1988}]%
        {sellis1988multiple}
\bibfield{author}{\bibinfo{person}{Timos~K Sellis}.}
  \bibinfo{year}{1988}\natexlab{}.
\newblock \showarticletitle{Multiple-query optimization}.
\newblock \bibinfo{journal}{\emph{TODS}} \bibinfo{volume}{13},
  \bibinfo{number}{1} (\bibinfo{year}{1988}), \bibinfo{pages}{23--52}.
\newblock


\bibitem[\protect\citeauthoryear{Shen, Han, and Wang}{Shen
  et~al\mbox{.}}{2014}]%
        {shen2014probabilistic}
\bibfield{author}{\bibinfo{person}{Wei Shen}, \bibinfo{person}{Jiawei Han},
  {and} \bibinfo{person}{Jianyong Wang}.} \bibinfo{year}{2014}\natexlab{}.
\newblock \showarticletitle{A probabilistic model for linking named entities in
  web text with heterogeneous information networks}. In
  \bibinfo{booktitle}{\emph{SIGMOD}}. \bibinfo{pages}{1199--1210}.
\newblock


\bibitem[\protect\citeauthoryear{Shi and Weninger}{Shi and Weninger}{2014}]%
        {DBLP:conf/icdm/ShiW14}
\bibfield{author}{\bibinfo{person}{Baoxu Shi} {and} \bibinfo{person}{Tim
  Weninger}.} \bibinfo{year}{2014}\natexlab{}.
\newblock \showarticletitle{Mining Interesting Meta-Paths from Complex
  Heterogeneous Information Networks}. In \bibinfo{booktitle}{\emph{{ICDM}}}.
  \bibinfo{pages}{488--495}.
\newblock


\bibitem[\protect\citeauthoryear{Shi, Hu, Zhao, and Yu}{Shi
  et~al\mbox{.}}{2019}]%
        {DBLP:journals/tkde/ShiHZY19}
\bibfield{author}{\bibinfo{person}{Chuan Shi}, \bibinfo{person}{Binbin Hu},
  \bibinfo{person}{Wayne~Xin Zhao}, {and} \bibinfo{person}{Philip~S. Yu}.}
  \bibinfo{year}{2019}\natexlab{}.
\newblock \showarticletitle{Heterogeneous Information Network Embedding for
  Recommendation}.
\newblock \bibinfo{journal}{\emph{TKDE}} \bibinfo{volume}{31},
  \bibinfo{number}{2} (\bibinfo{year}{2019}), \bibinfo{pages}{357--370}.
\newblock


\bibitem[\protect\citeauthoryear{Shi, Kong, Huang, Philip, and Wu}{Shi
  et~al\mbox{.}}{2014a}]%
        {Hetesim}
\bibfield{author}{\bibinfo{person}{Chuan Shi}, \bibinfo{person}{Xiangnan Kong},
  \bibinfo{person}{Yue Huang}, \bibinfo{person}{S~Yu Philip}, {and}
  \bibinfo{person}{Bin Wu}.} \bibinfo{year}{2014}\natexlab{a}.
\newblock \showarticletitle{Hetesim: A general framework for relevance measure
  in heterogeneous networks}.
\newblock \bibinfo{journal}{\emph{TKDE}} \bibinfo{volume}{26},
  \bibinfo{number}{10} (\bibinfo{year}{2014}), \bibinfo{pages}{2479--2492}.
\newblock


\bibitem[\protect\citeauthoryear{Shi, Kong, Yu, Xie, and Wu}{Shi
  et~al\mbox{.}}{2012}]%
        {shi2012}
\bibfield{author}{\bibinfo{person}{Chuan Shi}, \bibinfo{person}{Xiangnan Kong},
  \bibinfo{person}{Philip~S Yu}, \bibinfo{person}{Sihong Xie}, {and}
  \bibinfo{person}{Bin Wu}.} \bibinfo{year}{2012}\natexlab{}.
\newblock \showarticletitle{Relevance search in heterogeneous networks}. In
  \bibinfo{booktitle}{\emph{EDBT}}. \bibinfo{pages}{180--191}.
\newblock


\bibitem[\protect\citeauthoryear{Shi, Li, Yu, and Wu}{Shi
  et~al\mbox{.}}{2016a}]%
        {hrank}
\bibfield{author}{\bibinfo{person}{Chuan Shi}, \bibinfo{person}{Yitong Li},
  \bibinfo{person}{Philip~S. Yu}, {and} \bibinfo{person}{Bin Wu}.}
  \bibinfo{year}{2016}\natexlab{a}.
\newblock \showarticletitle{Constrained-meta-path-based ranking in
  heterogeneous information network}.
\newblock \bibinfo{journal}{\emph{Knowledge and Information Systems}}
  \bibinfo{volume}{49}, \bibinfo{number}{2} (\bibinfo{year}{2016}),
  \bibinfo{pages}{719--747}.
\newblock


\bibitem[\protect\citeauthoryear{Shi, Li, Zhang, Sun, and Philip}{Shi
  et~al\mbox{.}}{2016b}]%
        {HINsurvey}
\bibfield{author}{\bibinfo{person}{Chuan Shi}, \bibinfo{person}{Yitong Li},
  \bibinfo{person}{Jiawei Zhang}, \bibinfo{person}{Yizhou Sun}, {and}
  \bibinfo{person}{S~Yu Philip}.} \bibinfo{year}{2016}\natexlab{b}.
\newblock \showarticletitle{A survey of heterogeneous information network
  analysis}.
\newblock \bibinfo{journal}{\emph{TKDE}} \bibinfo{volume}{29},
  \bibinfo{number}{1} (\bibinfo{year}{2016}), \bibinfo{pages}{17--37}.
\newblock


\bibitem[\protect\citeauthoryear{Shi, Wang, Li, Yu, and Wu}{Shi
  et~al\mbox{.}}{2014b}]%
        {shi2014ranking}
\bibfield{author}{\bibinfo{person}{Chuan Shi}, \bibinfo{person}{Ran Wang},
  \bibinfo{person}{Yitong Li}, \bibinfo{person}{Philip~S Yu}, {and}
  \bibinfo{person}{Bin Wu}.} \bibinfo{year}{2014}\natexlab{b}.
\newblock \showarticletitle{Ranking-based clustering on general heterogeneous
  information networks by network projection}. In
  \bibinfo{booktitle}{\emph{CIKM}}. \bibinfo{pages}{699--708}.
\newblock


\bibitem[\protect\citeauthoryear{Shim, Scheuermann, and Vingralek}{Shim
  et~al\mbox{.}}{1999}]%
        {LNC-R-W3-U}
\bibfield{author}{\bibinfo{person}{Junho Shim}, \bibinfo{person}{Peter
  Scheuermann}, {and} \bibinfo{person}{Radek Vingralek}.}
  \bibinfo{year}{1999}\natexlab{}.
\newblock \showarticletitle{Proxy cache algorithms: Design, implementation, and
  performance}.
\newblock \bibinfo{journal}{\emph{TKDE}} \bibinfo{volume}{11},
  \bibinfo{number}{4} (\bibinfo{year}{1999}), \bibinfo{pages}{549--562}.
\newblock


\bibitem[\protect\citeauthoryear{{Shudong Jin} and {Bestavros}}{{Shudong Jin}
  and {Bestavros}}{2000}]%
        {Jin2000}
\bibfield{author}{\bibinfo{person}{{Shudong Jin}} {and} \bibinfo{person}{A.
  {Bestavros}}.} \bibinfo{year}{2000}\natexlab{}.
\newblock \showarticletitle{Popularity-aware greedy dual-size Web proxy caching
  algorithms}. In \bibinfo{booktitle}{\emph{ICDCS}}. \bibinfo{pages}{254--261}.
\newblock


\bibitem[\protect\citeauthoryear{Sommer, Boehm, Evfimievski, Reinwald, and
  Haas}{Sommer et~al\mbox{.}}{2019}]%
        {MNC}
\bibfield{author}{\bibinfo{person}{Johanna Sommer}, \bibinfo{person}{Matthias
  Boehm}, \bibinfo{person}{Alexandre~V. Evfimievski}, \bibinfo{person}{Berthold
  Reinwald}, {and} \bibinfo{person}{Peter~J. Haas}.}
  \bibinfo{year}{2019}\natexlab{}.
\newblock \showarticletitle{MNC: Structure-Exploiting Sparsity Estimation for
  Matrix Expressions}. In \bibinfo{booktitle}{\emph{SIGMOD}}.
  \bibinfo{pages}{1607--1623}.
\newblock


\bibitem[\protect\citeauthoryear{Starobinski and Tse}{Starobinski and
  Tse}{2001}]%
        {LRU_S}
\bibfield{author}{\bibinfo{person}{David Starobinski} {and}
  \bibinfo{person}{David Tse}.} \bibinfo{year}{2001}\natexlab{}.
\newblock \showarticletitle{Probabilistic methods for web caching}.
\newblock \bibinfo{journal}{\emph{Performance Evaluation}}
  \bibinfo{volume}{46}, \bibinfo{number}{2-3} (\bibinfo{year}{2001}),
  \bibinfo{pages}{125--137}.
\newblock


\bibitem[\protect\citeauthoryear{Sun, Aggarwal, and Han}{Sun
  et~al\mbox{.}}{2012a}]%
        {sun2012relation}
\bibfield{author}{\bibinfo{person}{Yizhou Sun}, \bibinfo{person}{Charu~C
  Aggarwal}, {and} \bibinfo{person}{Jiawei Han}.}
  \bibinfo{year}{2012}\natexlab{a}.
\newblock \showarticletitle{Relation strength-aware clustering of heterogeneous
  information networks with incomplete attributes}.
\newblock \bibinfo{journal}{\emph{arXiv preprint arXiv:1201.6563}}
  (\bibinfo{year}{2012}).
\newblock


\bibitem[\protect\citeauthoryear{Sun, Han, Yan, and Yu}{Sun
  et~al\mbox{.}}{2012b}]%
        {sun2012mining}
\bibfield{author}{\bibinfo{person}{Yizhou Sun}, \bibinfo{person}{Jiawei Han},
  \bibinfo{person}{Xifeng Yan}, {and} \bibinfo{person}{Philip~S Yu}.}
  \bibinfo{year}{2012}\natexlab{b}.
\newblock \showarticletitle{Mining knowledge from interconnected data: a
  heterogeneous information network analysis approach}.
\newblock \bibinfo{journal}{\emph{VLDB}} \bibinfo{volume}{5},
  \bibinfo{number}{12} (\bibinfo{year}{2012}), \bibinfo{pages}{2022--2023}.
\newblock


\bibitem[\protect\citeauthoryear{Sun, Han, Yan, Yu, and Wu}{Sun
  et~al\mbox{.}}{2011}]%
        {sun11}
\bibfield{author}{\bibinfo{person}{Yizhou Sun}, \bibinfo{person}{Jiawei Han},
  \bibinfo{person}{Xifeng Yan}, \bibinfo{person}{Philip~S. Yu}, {and}
  \bibinfo{person}{Tianyi Wu}.} \bibinfo{year}{2011}\natexlab{}.
\newblock \showarticletitle{{PathSim}: Meta Path-Based Top-$k$ Similarity
  Search in Heterogeneous Information Networks}.
\newblock \bibinfo{journal}{\emph{{PVLDB}}} \bibinfo{volume}{4},
  \bibinfo{number}{11} (\bibinfo{year}{2011}), \bibinfo{pages}{992--1003}.
\newblock


\bibitem[\protect\citeauthoryear{Sun, Han, Zhao, Yin, Cheng, and Wu}{Sun
  et~al\mbox{.}}{2009a}]%
        {sun2009rankclus}
\bibfield{author}{\bibinfo{person}{Yizhou Sun}, \bibinfo{person}{Jiawei Han},
  \bibinfo{person}{Peixiang Zhao}, \bibinfo{person}{Zhijun Yin},
  \bibinfo{person}{Hong Cheng}, {and} \bibinfo{person}{Tianyi Wu}.}
  \bibinfo{year}{2009}\natexlab{a}.
\newblock \showarticletitle{Rankclus: integrating clustering with ranking for
  heterogeneous information network analysis}. In
  \bibinfo{booktitle}{\emph{EDBT}}. \bibinfo{pages}{565--576}.
\newblock


\bibitem[\protect\citeauthoryear{Sun, Norick, Han, Yan, Yu, and Yu}{Sun
  et~al\mbox{.}}{2013}]%
        {sun2013pathselclus}
\bibfield{author}{\bibinfo{person}{Yizhou Sun}, \bibinfo{person}{Brandon
  Norick}, \bibinfo{person}{Jiawei Han}, \bibinfo{person}{Xifeng Yan},
  \bibinfo{person}{Philip~S Yu}, {and} \bibinfo{person}{Xiao Yu}.}
  \bibinfo{year}{2013}\natexlab{}.
\newblock \showarticletitle{Pathselclus: Integrating meta-path selection with
  user-guided object clustering in heterogeneous information networks}.
\newblock \bibinfo{journal}{\emph{TKDD}} \bibinfo{volume}{7},
  \bibinfo{number}{3} (\bibinfo{year}{2013}), \bibinfo{pages}{1--23}.
\newblock


\bibitem[\protect\citeauthoryear{Sun, Yu, and Han}{Sun et~al\mbox{.}}{2009b}]%
        {sun2009ranking}
\bibfield{author}{\bibinfo{person}{Yizhou Sun}, \bibinfo{person}{Yintao Yu},
  {and} \bibinfo{person}{Jiawei Han}.} \bibinfo{year}{2009}\natexlab{b}.
\newblock \showarticletitle{Ranking-based clustering of heterogeneous
  information networks with star network schema}. In
  \bibinfo{booktitle}{\emph{SIGKDD}}. \bibinfo{pages}{797--806}.
\newblock


\bibitem[\protect\citeauthoryear{Szpankowski}{Szpankowski}{1993}]%
        {generalised_suffix_tree_1}
\bibfield{author}{\bibinfo{person}{Wojciech Szpankowski}.}
  \bibinfo{year}{1993}\natexlab{}.
\newblock \showarticletitle{A generalized suffix tree and its (un) expected
  asymptotic behaviors}.
\newblock \bibinfo{journal}{\emph{SICOMP}} \bibinfo{volume}{22},
  \bibinfo{number}{6} (\bibinfo{year}{1993}), \bibinfo{pages}{1176--1198}.
\newblock


\bibitem[\protect\citeauthoryear{Tang, Zhang, Yao, Li, Zhang, and Su}{Tang
  et~al\mbox{.}}{2008}]%
        {aminer}
\bibfield{author}{\bibinfo{person}{Jie Tang}, \bibinfo{person}{Jing Zhang},
  \bibinfo{person}{Limin Yao}, \bibinfo{person}{Juanzi Li}, \bibinfo{person}{Li
  Zhang}, {and} \bibinfo{person}{Zhong Su}.} \bibinfo{year}{2008}\natexlab{}.
\newblock \showarticletitle{Arnet{M}iner: {E}xtraction and {M}ining of
  {A}cademic {S}ocial {N}etworks}. In \bibinfo{booktitle}{\emph{SIGKDD}}.
  \bibinfo{pages}{990--998}.
\newblock


\bibitem[\protect\citeauthoryear{Ukkonen}{Ukkonen}{1995}]%
        {ukkonen1995}
\bibfield{author}{\bibinfo{person}{Esko Ukkonen}.}
  \bibinfo{year}{1995}\natexlab{}.
\newblock \showarticletitle{On-line construction of suffix trees}.
\newblock \bibinfo{journal}{\emph{Algorithmica}} \bibinfo{volume}{14},
  \bibinfo{number}{3} (\bibinfo{year}{1995}), \bibinfo{pages}{249--260}.
\newblock


\bibitem[\protect\citeauthoryear{Wang, Wang, Zhao, Li, Jian, Chen, and
  Song}{Wang et~al\mbox{.}}{2020}]%
        {wang2020howsim}
\bibfield{author}{\bibinfo{person}{Yue Wang}, \bibinfo{person}{Zhe Wang},
  \bibinfo{person}{Ziyuan Zhao}, \bibinfo{person}{Zijian Li},
  \bibinfo{person}{Xun Jian}, \bibinfo{person}{Lei Chen}, {and}
  \bibinfo{person}{Jianchun Song}.} \bibinfo{year}{2020}\natexlab{}.
\newblock \showarticletitle{HowSim: A General and Effective Similarity Measure
  on Heterogeneous Information Networks}. In \bibinfo{booktitle}{\emph{ICDE}}.
  \bibinfo{pages}{1954--1957}.
\newblock


\bibitem[\protect\citeauthoryear{Webber}{Webber}{2012}]%
        {webber2012}
\bibfield{author}{\bibinfo{person}{Jim Webber}.}
  \bibinfo{year}{2012}\natexlab{}.
\newblock \showarticletitle{A programmatic introduction to neo4j}. In
  \bibinfo{booktitle}{\emph{SPLASH}}. \bibinfo{pages}{217--218}.
\newblock


\bibitem[\protect\citeauthoryear{Weiss, Karras, and Bernstein}{Weiss
  et~al\mbox{.}}{2008}]%
        {hexastore}
\bibfield{author}{\bibinfo{person}{Cathrin Weiss}, \bibinfo{person}{Panagiotis
  Karras}, {and} \bibinfo{person}{Abraham Bernstein}.}
  \bibinfo{year}{2008}\natexlab{}.
\newblock \showarticletitle{Hexastore: sextuple indexing for semantic web data
  management}.
\newblock \bibinfo{journal}{\emph{VLDB}} \bibinfo{volume}{1},
  \bibinfo{number}{1} (\bibinfo{year}{2008}), \bibinfo{pages}{1008--1019}.
\newblock


\bibitem[\protect\citeauthoryear{Xie, Xu, Chen, Liu, and Zheng}{Xie
  et~al\mbox{.}}{2021}]%
        {xie2021sequential}
\bibfield{author}{\bibinfo{person}{Tao Xie}, \bibinfo{person}{Yangjun Xu},
  \bibinfo{person}{Liang Chen}, \bibinfo{person}{Yang Liu}, {and}
  \bibinfo{person}{Zibin Zheng}.} \bibinfo{year}{2021}\natexlab{}.
\newblock \showarticletitle{Sequential Recommendation on Dynamic Heterogeneous
  Information Network}. In \bibinfo{booktitle}{\emph{ICDE}}.
  \bibinfo{pages}{2105--2110}.
\newblock


\bibitem[\protect\citeauthoryear{Y.~Xiong}{Y.~Xiong}{2014}]%
        {joinsim}
\bibfield{author}{\bibinfo{person}{P.~Yu Y.~Xiong, Y.~Zhu}.}
  \bibinfo{year}{2014}\natexlab{}.
\newblock \showarticletitle{Top-k Similarity Join in Heterogeneous Information
  Networks}.
\newblock \bibinfo{journal}{\emph{TKDE}} \bibinfo{volume}{27},
  \bibinfo{number}{6} (\bibinfo{year}{2014}), \bibinfo{pages}{1710--1723}.
\newblock


\bibitem[\protect\citeauthoryear{Yang, Fang, Lin, and Zhang}{Yang
  et~al\mbox{.}}{2020}]%
        {yang2020effective}
\bibfield{author}{\bibinfo{person}{Yixing Yang}, \bibinfo{person}{Yixiang
  Fang}, \bibinfo{person}{Xuemin Lin}, {and} \bibinfo{person}{Wenjie Zhang}.}
  \bibinfo{year}{2020}\natexlab{}.
\newblock \showarticletitle{Effective and efficient truss computation over
  large heterogeneous information networks}. In
  \bibinfo{booktitle}{\emph{ICDE}}. \bibinfo{pages}{901--912}.
\newblock


\bibitem[\protect\citeauthoryear{Young}{Young}{1994}]%
        {GD}
\bibfield{author}{\bibinfo{person}{Neal Young}.}
  \bibinfo{year}{1994}\natexlab{}.
\newblock \showarticletitle{Thek-server dual and loose competitiveness for
  paging}.
\newblock \bibinfo{journal}{\emph{Algorithmica}} \bibinfo{volume}{11},
  \bibinfo{number}{6} (\bibinfo{year}{1994}), \bibinfo{pages}{525--541}.
\newblock


\bibitem[\protect\citeauthoryear{Yu, Tang, Aref, Malluhi, Abbas, and
  Ouzzani}{Yu et~al\mbox{.}}{2017}]%
        {MatFast}
\bibfield{author}{\bibinfo{person}{Yongyang Yu}, \bibinfo{person}{Mingjie
  Tang}, \bibinfo{person}{{Walid G.} Aref}, \bibinfo{person}{{Qutaibah M.}
  Malluhi}, \bibinfo{person}{Mostafa Abbas}, {and} \bibinfo{person}{Mourad
  Ouzzani}.} \bibinfo{year}{2017}\natexlab{}.
\newblock \showarticletitle{In-memory distributed matrix computation processing
  \& optimization}. In \bibinfo{booktitle}{\emph{ICDE}}.
  \bibinfo{pages}{1047--1058}.
\newblock


\bibitem[\protect\citeauthoryear{Zhou and Liu}{Zhou and Liu}{2013}]%
        {zhou2013social}
\bibfield{author}{\bibinfo{person}{Yang Zhou} {and} \bibinfo{person}{Ling
  Liu}.} \bibinfo{year}{2013}\natexlab{}.
\newblock \showarticletitle{Social influence based clustering of heterogeneous
  information networks}. In \bibinfo{booktitle}{\emph{SIGKDD}}.
  \bibinfo{pages}{338--346}.
\newblock


\end{thebibliography}

\end{document}